\documentclass[twocolumn]{revtex4-1}
\usepackage{bm}
\usepackage{comment}
\usepackage{amsmath}
\usepackage{amsfonts}
\usepackage{amssymb}

\date{\today}
\usepackage[pdftex]{graphicx}

\usepackage{wrapfig}
\usepackage{here}

\begin{document}

\title{
Thermal Hall Conductivity in Superconducting Phase on Kagome Lattice\textit{}
}

\author{Shoma Iimura$^{1}$} 
\email{s.iimura.787@ms.saitama-u.ac.jp}
\author{Yoshiki Imai$^2$}

\affiliation{
$^1$Department of Physics, Saitama University, Shimo-Okubo 255, Sakura-ku, Saitama-shi,  338-8570 Saitama, Japan\\
$^2$Department of Applied Physics, Okayama University of Science, 1-1 Ridaicho, Kita-ku, 700-0005 Okayama, Japan
} 

\begin{abstract}
{\footnotesize
Motivated by a previous ``$sd^2$-graphene'' study, the pairing symmetry in the superconducting state and the thermal Hall conductivity are investigated by a self-consistent Bogoliubov--de Gennes approach on the kagome lattice with intrinsic spin-orbit coupling near van Hove fillings. 
While the topologically trivial state with broken time-reversal symmetry appears in the absence of spin-orbit coupling, the highest flat band becomes dispersive with a hexagonal symmetry due to spin-orbit coupling, which leads to a topological superconducting state. Since the thermal Hall conductivity in the low-temperature limit is associated with the topological property of time-reversal symmetry breaking superconductors, we study its temperature dependence near van Hove fillings.  
In particular, the pairing symmetry in the highest flat band is sensitive to the amplitudes of spin-orbit coupling and the attractive interaction, which is reflected remarkably in the thermal Hall conductivity. 
The obtained result may enable us to investigate the stable superconducting state on the kagome lattice.}
\end{abstract}

\maketitle

\section{Introduction}
Chiral superconductivity has attracted much interest as a hot topic in condensed matter physics. The pairing in the superconducting state breaks time-reversal symmetry (TRS) and has the nontrivial topology with interesting properties \cite{Topological_insulators_and_superconductors,Kallin,Hasan}.

Experimentally, the chiral $p$-wave is observed in the $A$-phase of the spin-triplet superfluid $^3 \mathrm{He}$~\cite{he3}, and the transition metal oxide $\mathrm{Sr}_2 \mathrm{Ru} \mathrm{O}_4$ is one of the prime candidates~\cite{SRO_1,SRO_2,SRO_3}. On the other hand, the spin-singlet $d_{x^2 - y^2} + id_{xy}$ state ($d+id$ state) is also another chiral superconducting state. Although it has not been confirmed experimentally, there are some potential candidate materials such as heavily doped graphene and SrPtAs~\cite{SrPtAs}. 

Doped graphene corresponding to $3/8$ or $5/8$ filling has a density of state (DOS) with logarithmic divergence, which results from the van Hove singularities (VHSs) originating from three inequivalent saddle points where the Fermi surface has a hexagonal geometry with perfect nesting. The renormalization group studies for the Hubbard model~\cite{PRG_Graphen, FRG_Graphen1,FRG_Graphen2} indicate that the $d+id$ state is favored in heavily doped graphene. 
On the basis of the mean-field approximation, it is also confirmed that the attractive interaction between nearest-neighbor (NN) sites or next nearest-neighbor (NNN) sites leads to the $d+id$ state in the vicinity of the van Hove (VH) fillings~\cite{chiral_d_Graphen}. 

\begin{figure}[t]
\centering
\includegraphics[angle=0,width=8.4cm,height=4.0cm]{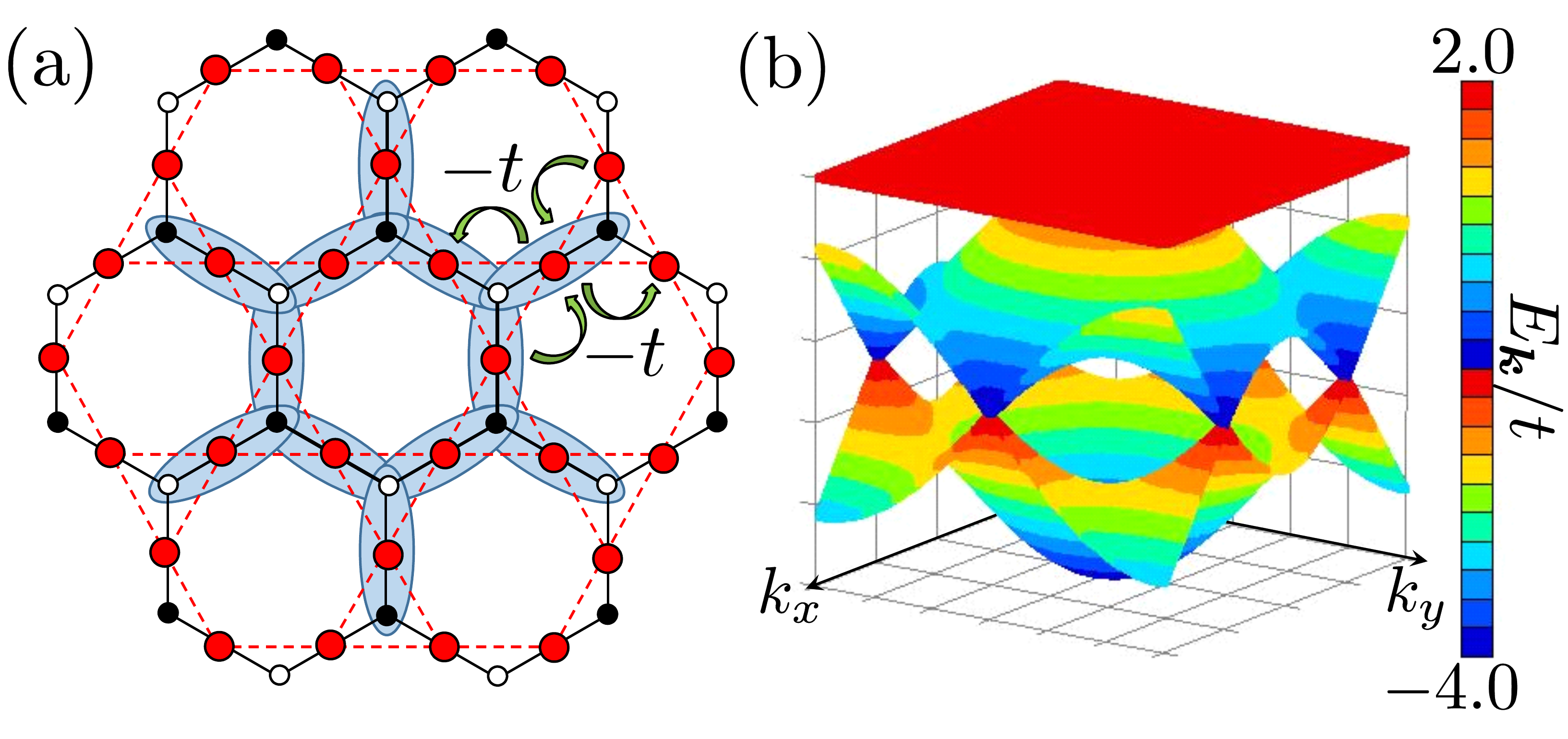}
\caption{(Color online) (a) Schematic diagram of the $sd^2$-graphene. The low-energy properties of the bond-centered $\sigma$-states (red points) are effectively described by NN hopping ($-t$) on the kagome lattice (red dashed lines). (b) Energy dispersion of the kagome lattice in the absence of spin-orbit coupling.}
\label{kagome}
\end{figure}

Graphene is composed of $sp^2$ hybridization in carbon, and its low-energy electronic structure is well described by a tight-binding model with the hopping between NN sites on a hexagonal lattice\cite{Graphen}. However, some honeycomb lattice structures consisting of transition metal atoms are composed of ``$sd^2$-hybridization'' and called ``$sd^2$-graphene''~\cite{sd2_Graphen}, where the electron transfers between bond-centered $\sigma$ states on a hexagonal lattice. Therefore, the low-energy property is effectively described by single-orbital hopping between NN sites on the kagome lattice, shown in Fig. \ref{kagome}(a). One of the candidates is a hexagonal $\mathrm{W}$ lattice epitaxially grown on a semiconductor surface\cite{sd2_Graphen}. Transition metal atoms may lead to a large amplitude of the spin-orbit coupling (SOC) compared with graphene. 

The single orbital tight-binding model on the kagome lattice with NN hopping has two dispersive bands and a completely dispersionless flat band, which is depicted in Fig. \ref{kagome}(b). The former bands are the same as those in the honeycomb lattice and are hereafter called ``honeycomb bands'', which include VHSs resulting from the saddle points. 
The latter band is called ``highest band''.  
The spin systems on the kagome lattice show interesting behavior such as the quantum spin liquid in the ground state\cite{kagome1,kagome2,kagome3} due to the characteristic lattice geometry. However, there is also extensive interest in itinerant electron systems on the kagome lattice. In the vicinity of VH fillings on the kagome lattice, short-range repulsive interactions may generate the $d+id$-wave pairing on 3rd NN bonds obtained by the variational cluster approach\cite{VCA_kagome} and the renormalization group studies\cite{Wang2013,Kiesel_1,Kiesel_2}.
 
In addition, the intrinsic SOC generates rich phases. In the Kane--Mele topological insulator on  the honeycomb lattice\cite{Kane_Mele,Kane_Mele2}, the SOC connects between NNN sites because the two sides are asymmetric about NNN bonds, and the gradient of electrostatic potential appears. However, in the kagome lattice, there is already such an asymmetry about NN bonds. 
The singular-mode functional renormalization group study expected that the strong SOC may generate various orders such as magnetism and superconductivity\cite{SMFRG_kagome} in the honeycomb bands. 
On the other hand, the dispersionless flat band on the kagome lattice is very sensitive to the presence of the SOC, and generates a large DOS, which may give rise to the superconducting state with a high transition temperature.  

On the other hand, thermal Hall conductivity is suitable for investigating the topological property of TRS breaking in superconductors\cite{Thermal_Hall_conductivity}. In the low-temperature limit, thermal Hall conductivity is proportional to temperature and it may also be possible to detect the chiral superconducting state. 

In this paper, we focus on the superconducting state in the vicinity of the highest band on the kagome lattice. We calculate thermal Hall conductivity to distinguish the topological aspect of the superconducting state by using the lattice model, and discuss the interplay between the SOC and superconductivity. 

The paper is organized as follows. We construct the effective Hamiltonian in Sec. 2. In Sec. 3, we show the normal phase properties induced by the SOC. After that, we show the phase diagram of superconducting states, the temperature dependence of order parameters, and the thermal Hall conductivities in Sec. 4. Conclusions are given in Sec. 5.

\section{Model}
\label{MODEL}
\begin{figure}[t]
\begin{center}
\includegraphics[width=8.4cm]{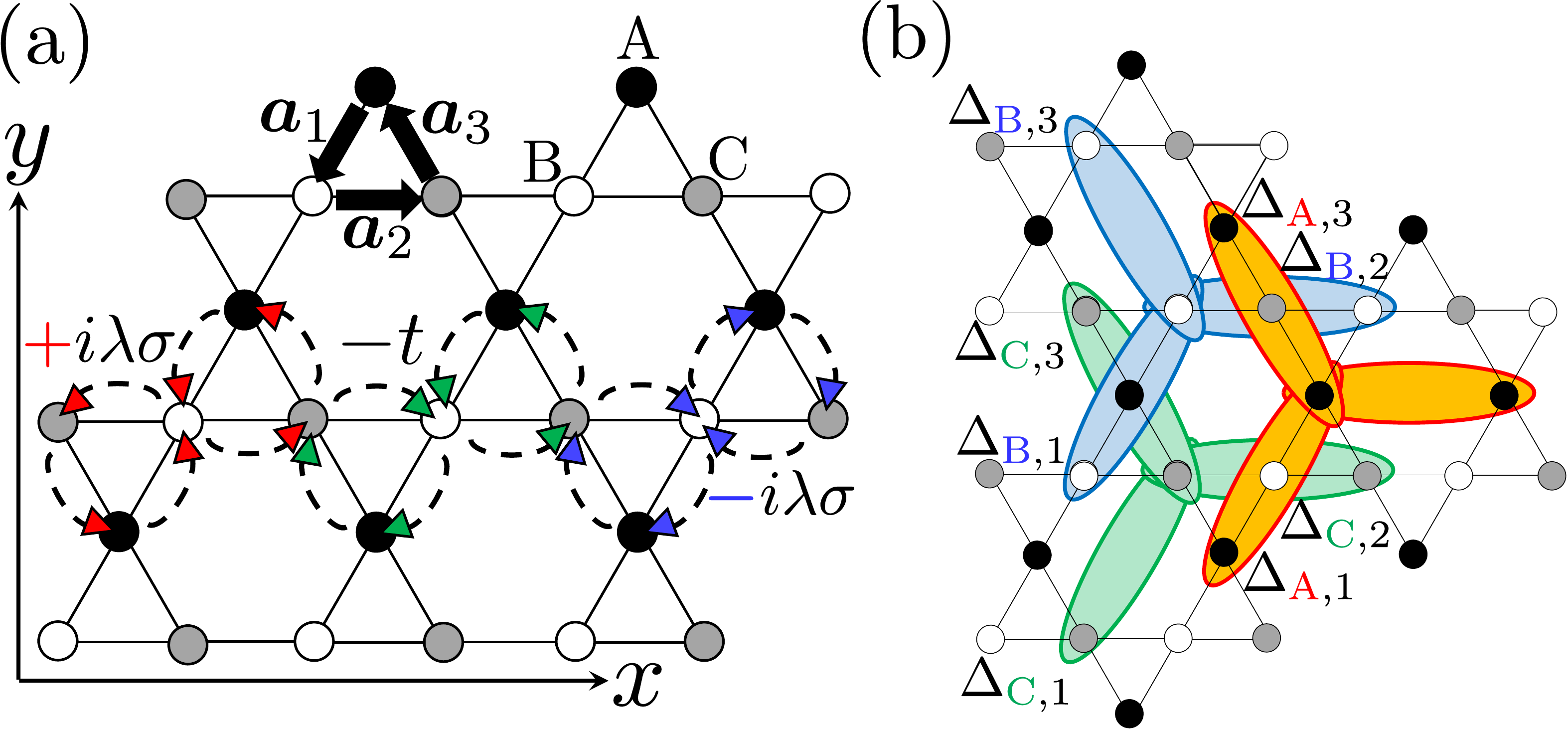}
\caption{(Color online) (a) Structure of kagome lattice. $t$ and $\lambda$ stand for the hopping integral and spin-orbit coupling between NN sites, respectively. $\bm{a}_i~(i=1,2,3)$ represents the displacement vector between NN sites. (b) Attractive interaction between 3rd NN sites. The ellipses correspond to the superconducting pairing $\Delta_{M,i}$ between the same sublattices. }
\label{fig_model}
\end{center}
\end{figure}

In this section, we introduce the model Hamiltonian in order to discuss the superconducting state on the kagome lattice. Figure \ref{fig_model}(a) shows the lattice structure, which is composed of A, B, and C sublattices. $\bm{a}_1$, $\bm{a}_2$, and $\bm{a}_3$ are the displacement vectors between NN sites. $t$ and $\lambda$ denote the amplitude of the hopping and the SOC between NN sites respectively. 

Since short-range repulsive interactions generate the $d+id$-wave pairing between the 3rd NN sites near the VHS in the upper honeycomb band\cite{VCA_kagome,Wang2013}, we introduce the attractive interaction between the same sublattices. The schematic picture is drawn in Fig. 2(b).

The corresponding Hamiltonian is written as
\begin{align}
H = H_{h} + H_{so} + H_a, 
\end{align}
with
\begin{align}
\label{Hamiltonian}
H_{h} &= -t \sum_{\left< i,j \right> , \sigma} \left( c_{i\sigma}^\dagger c_{j\sigma}^{ } + \mathrm{H}.c. \right) - \mu \sum_{i,\sigma} n_{i\sigma}^{ }, \nonumber \\
H_{so} &= i\lambda \sum_{\left< i,j\right> , \sigma} \sigma \nu_{ij} c_{i\sigma}^\dagger c_{j\sigma}^{ } + \mathrm{H}.c., \\
H_a &= -U \sum_{\left\{ i,j \right\} ,\sigma} n_{i\sigma}^{ } n_{j\bar{\sigma}}^{ }, \nonumber  
\end{align}
where $i$ (and $j$) is the site index including the sublattice index $M(=A,B,C)$. Then $c_{i\sigma}^\dagger (c_{i\sigma}^{ })$ denotes the creation (annihilation) operator for the electron on site $i$ with spin $\sigma = \uparrow , \downarrow$. $n_{i\sigma} = c_{i\sigma}^\dagger c_{i\sigma}^{ }$ is the particle number operator. $\left< i,j\right>$ $(\left\{ i,j \right\})$ stands for the summation for NN (3rd NN) bonds.

$H_{h}$ represents the hopping term between NN sites with the amplitude $t$ and the chemical potential $\mu$. $H_{so}$ represents the intrinsic SOC between NN sites. $\nu_{ij}$ is a factor corresponding to the hopping direction, which is $-1(+1)$ for the clockwise (counterclockwise) direction. $H_a$ represents the attractive interaction on 3rd NN bonds with the amplitude $-U$. 
We introduce the BCS-type mean-field approximation to decouple the attractive interaction term assuming the spin-singlet channel. 
Then the interaction term $H_a$ is rewritten as
\begin{align}
H_a\approx H_a^{MF} = -U\sum_{\left\{ i,j \right\} , \sigma} \sigma \Delta_{ij} c_{i,\sigma}^\dagger c_{j,-\sigma}^\dagger + \mathrm{H}.c. + const., 
\end{align}
where $\Delta_{ij}$ is the spin-singlet superconducting order parameter defined as 
\begin{align}
\label{gap_function}
\Delta_{M,l} \equiv \frac{1}{2}\left[ \left< c_{j,\downarrow}^{ } c_{i,\uparrow}^{ } \right> - \left< c_{j,\uparrow}^{ } c_{i,\downarrow}^{ } \right> \right], 
\end{align}
where $M$ and $l$ stand for the sublattice and displacement indices, respectively.

Then, we obtain  the following Bogoliubov--de Gennes (BdG) Hamiltonian in momentum space, yielding the 3rd NN spin-singlet pairing on the kagome lattice,
\begin{align}
\label{BdG}
H_{\mathrm{BdG}} = \frac{1}{2} \sum_{\bm{k},\sigma}  \Psi_{\bm{k},\sigma}^\dagger 
\begin{pmatrix}
\hat{\xi}_\sigma(\bm{k}) & \hat{\Delta}_\sigma(\bm{k}) \\
\hat{\Delta}_\sigma(\bm{k})^\dagger & -\hat{\xi}_{-\sigma}(-\bm{k}) 
\end{pmatrix}
\Psi_{\bm{k},\sigma}^{ },
\end{align}
where the operator $\Psi_{\bm{k},\sigma}^\dagger = \left( \psi_{\bm{k},\sigma}^\dagger , \psi_{-\bm{k},-\sigma}^{ } \right)$ is a six-component Nambu spinor representation with $\psi_{\bm{k},\sigma} = \left( c_{\bm{k},A,\sigma},c_{\bm{k},B,\sigma},c_{\bm{k},C,\sigma} \right)^T$. $c_{\bm{k},M,\sigma}$ denotes the Fourier component of the electron annihilation operator $c_{i,M,\sigma}$.

The diagonal block term $\hat{\xi}_\sigma(\bm{k})$, which describes the energy dispersion resulting from the kinetic term, is given by
\begin{align}
\label{kinetic_term}
\hat{\xi}_\sigma(\bm{k}) = 
\begin{pmatrix}
-\mu & \varepsilon_{AB,\sigma}(\bm{k}) & \varepsilon_{AC,\sigma}(\bm{k})\\
\varepsilon_{BA,\sigma}(\bm{k}) & -\mu & \varepsilon_{BC,\sigma}(\bm{k})\\
\varepsilon_{CA,\sigma}(\bm{k}) & \varepsilon_{CB,\sigma}(\bm{k}) & -\mu
\end{pmatrix},
\end{align}
with
\begin{align}
\left\{ \begin{array}{c}
\varepsilon_{AB,\sigma}(\bm{k}) = \left( \varepsilon_{BA,\sigma}(\bm{k})\right)^* = -2(t+i\lambda \sigma) \mathrm{cos}\left(\bm{k}\cdot\bm{a}_1\right) \\
\varepsilon_{BC,\sigma}(\bm{k}) = \left(\varepsilon_{CB,\sigma}(\bm{k})\right)^* = -2(t+i\lambda\sigma)\mathrm{cos}\left(\bm{k}\cdot\bm{a}_2\right) \\
\varepsilon_{CA,\sigma}(\bm{k}) = \left( \varepsilon_{AC,\sigma}(\bm{k})\right)^* = -2(t+i\lambda \sigma) \mathrm{cos}\left(\bm{k}\cdot\bm{a}_3\right) 
\end{array} \right..
\end{align}

Moreover, the off-diagonal block term $\hat{\Delta}_\sigma(\bm{k})$ in $H_{\mathrm{BdG}}$, which represents the superconducting order parameter, is given by
\begin{align}
\label{gap_SC}
\hat{\Delta}_\sigma (\bm{k}) = 
\begin{pmatrix}
\Delta_{A,\sigma}(\bm{k}) & 0 & 0\\
0 & \Delta_{B,\sigma}(\bm{k}) & 0\\
0 & 0 & \Delta_{C,\sigma}(\bm{k}) 
\end{pmatrix}, 
\end{align}
with 
\begin{align}
\Delta_{M,\sigma}(\bm{k}) = 2U\sigma\sum_{i=1}^3 \Delta_{M,i} \mathrm{cos}(2\bm{k}\cdot \bm{a}_i), 
\end{align}
where the symbols $\Delta_{M,i}$ correspond to the order parameter in Eq. (\ref{gap_function}). 

Hereafter, we will define the lattice constants $2\left|\bm{a}_i\right| = a$ and take the physical constants $\hbar$, $e$, $k_{\mathrm{B}}$ and $a$ as unity. 

\section{Normal Phase}
First, we discuss physical properties in the normal phase before discussing the superconducting phase.  
The band structure, Fermi surface, and topological property are investigated for $U=0$ in this section. 

\subsection{Energy dispersion and Fermi surface}
\begin{figure}[t]
	\begin{center}
		\begin{tabular}{c}
			\begin{minipage}{7.0cm}
			\begin{center}
		\includegraphics[width=7.0cm]{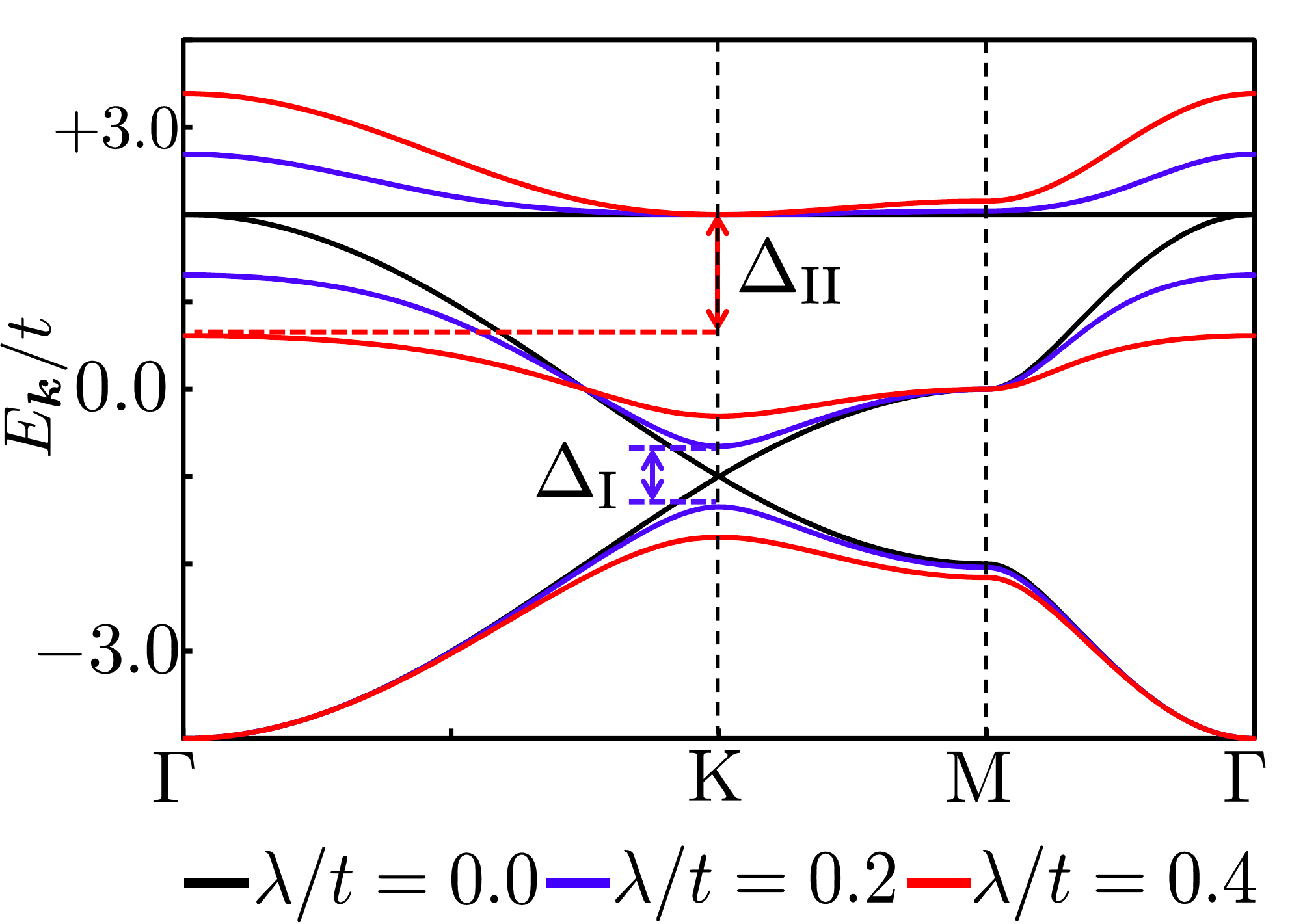}
		\caption{(Color online) Energy dispersions for several choices of $\lambda$. 
			The black solid, blue solid, and red solid lines correspond to $\lambda/t = 0.0, 0.2$, and $0.4$ respectively.}
		\label{band_SOC}
			\end{center}
			\end{minipage}\\
			\vspace{1.0cm}\\
			\begin{minipage}{7.0cm}
			\begin{center}
			\includegraphics[angle=90,width=7.0cm]{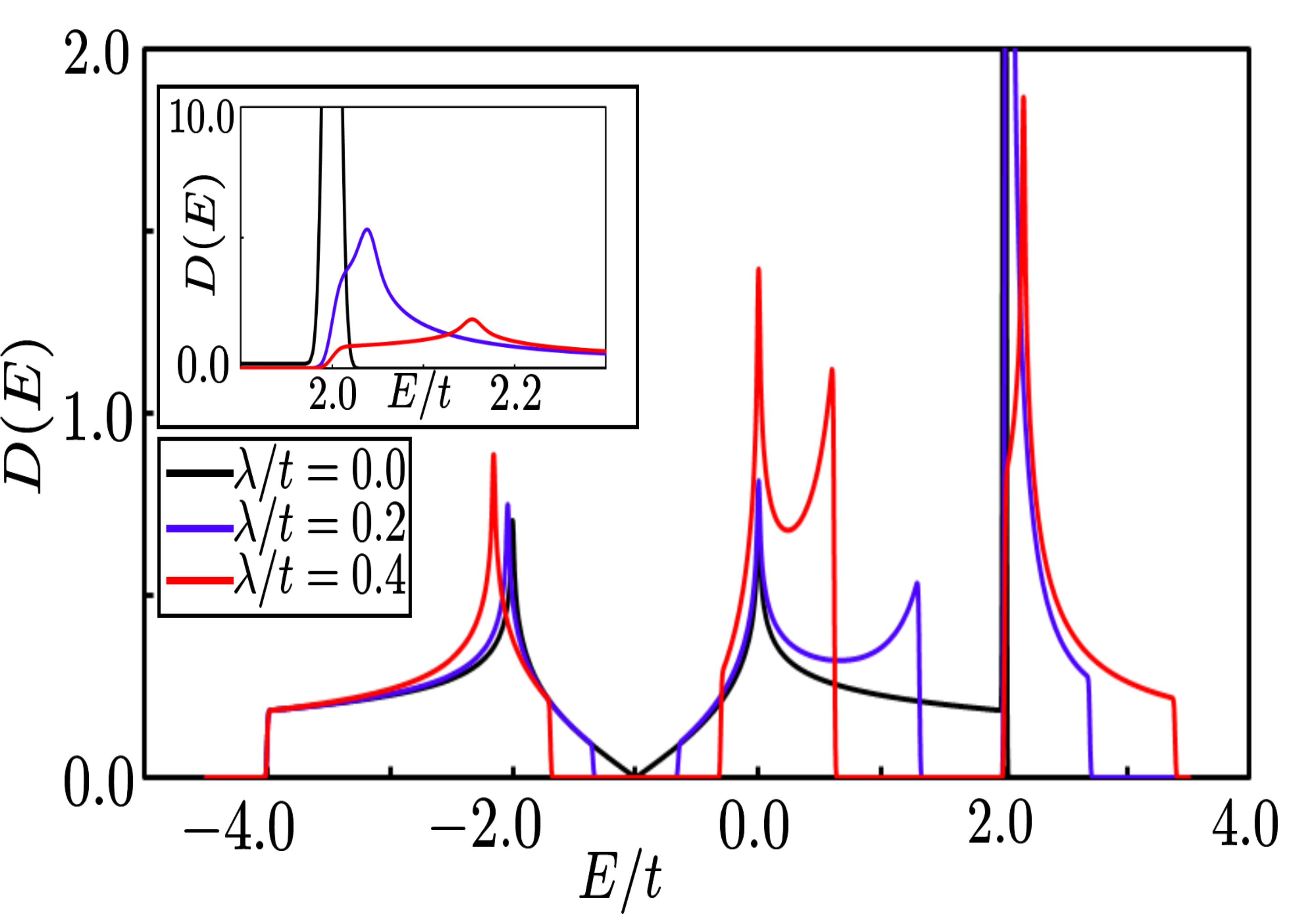}
		\caption{(Color online) Density of states for several choices of $\lambda$. The black, red, and blue lines correspond to $\lambda/t = 0.0,0.2$, and $0.4$, respectively. The inset shows the magnification of the DOS around $\mu/t=2$.}
			\label{DOS}
			\end{center}
			\end{minipage}
		\end{tabular}
		\vspace{-0.8cm}
	\end{center}
\end{figure}

Figure \ref{band_SOC} shows the energy dispersion $E_{\bm k}$ for several choices of $\lambda$ obtained from the diagonalization of $\hat{\xi}_\sigma(\bm{k})$ in Eq. (\ref{kinetic_term}).  
In the absence of the SOC, there exist three bands, which consist of  the dispersive honeycomb bands and the highest dispersionless flat band. 
The SOC generates the fully opened gap $\Delta_{\mathrm{I}}$ ($\Delta_{\mathrm{II}}$), which is defined by the amplitude between the top of the lower honeycomb (upper honeycomb) band and the bottom of the upper honeycomb (highest) band. 

The highest band switches from dispersionless to dispersive owing to the introduction of the SOC, whose amplitude at the K point is independent of $\lambda$ ($E_{\bm k}=2t$), and the M point becomes the saddle point for $\lambda\ne 0$.  
The width of the upper honeycomb band is reduced strongly with increasing $\lambda$ and becomes completely flat at $\lambda=\lambda_c (\equiv  t/\sqrt{3} \approx  0.577t)$.  
The amplitudes of energy gaps are given by $\Delta_{\mathrm{I}}=\Delta_{\mathrm{II}}=2\sqrt{3}\lambda$ for $\lambda \leq \lambda_c$.
Although further increase in $\lambda$ leads to  $\Delta_{\mathrm{I}}=\Delta_{\mathrm{II}}=3t - \sqrt{3} \lambda $ for $\lambda \geq \lambda_c$, we focus on the case of $\lambda<\lambda_c$ in this study.  

Figure \ref{DOS} shows the DOS for several choices of $\lambda$. 
There are three characteristic peaks resulting from VHSs at $E=-2\sqrt{t^2+\lambda^2}$, $0$, and $+2\sqrt{t^2+\lambda^2}$. 
While peak positions in the lower honeycomb band and highest band shift to lower- and higher-energy sides with increasing $\lambda$,  that in the upper honeycomb band is independent of the amplitude of $\lambda$. 
Note that $\mu=\mu_1\equiv -2\sqrt{t^2+\lambda^2}$ and $\mu_2\equiv 0$ correspond to the particle number $\langle n \rangle = 1/4$ and $5/12$,  respectively. 
With increasing $\lambda$, the amplitudes of the DOS at VH fillings increase at $\left< n \right> = 1/4$ and $5/12$, as shown in Fig. \ref{DOS}. The latter case ($\left< n \right> = 5/12$) is more remarkable since the SOC reduces strongly the width of the upper honeycomb band. 
 
Moreover, $\mu=\mu_3\equiv +2\sqrt{t^2+\lambda^2}$ corresponds to $\langle n \rangle = 3/4$ except for $\lambda =0$, because in the absence of $\lambda$, the highest band becomes completely flat. 
The SOC turns the dispersionless flat band into a dispersive one. 
Therefore, the amplitude of the peak at $E=\mu_3$ is more sensitive to the amplitude of $\lambda$ than those at $E=\mu_1$ and $E=\mu_2$. 

\begin{figure}[t]
	\begin{center}
		\includegraphics[angle=0,width=7.0cm]{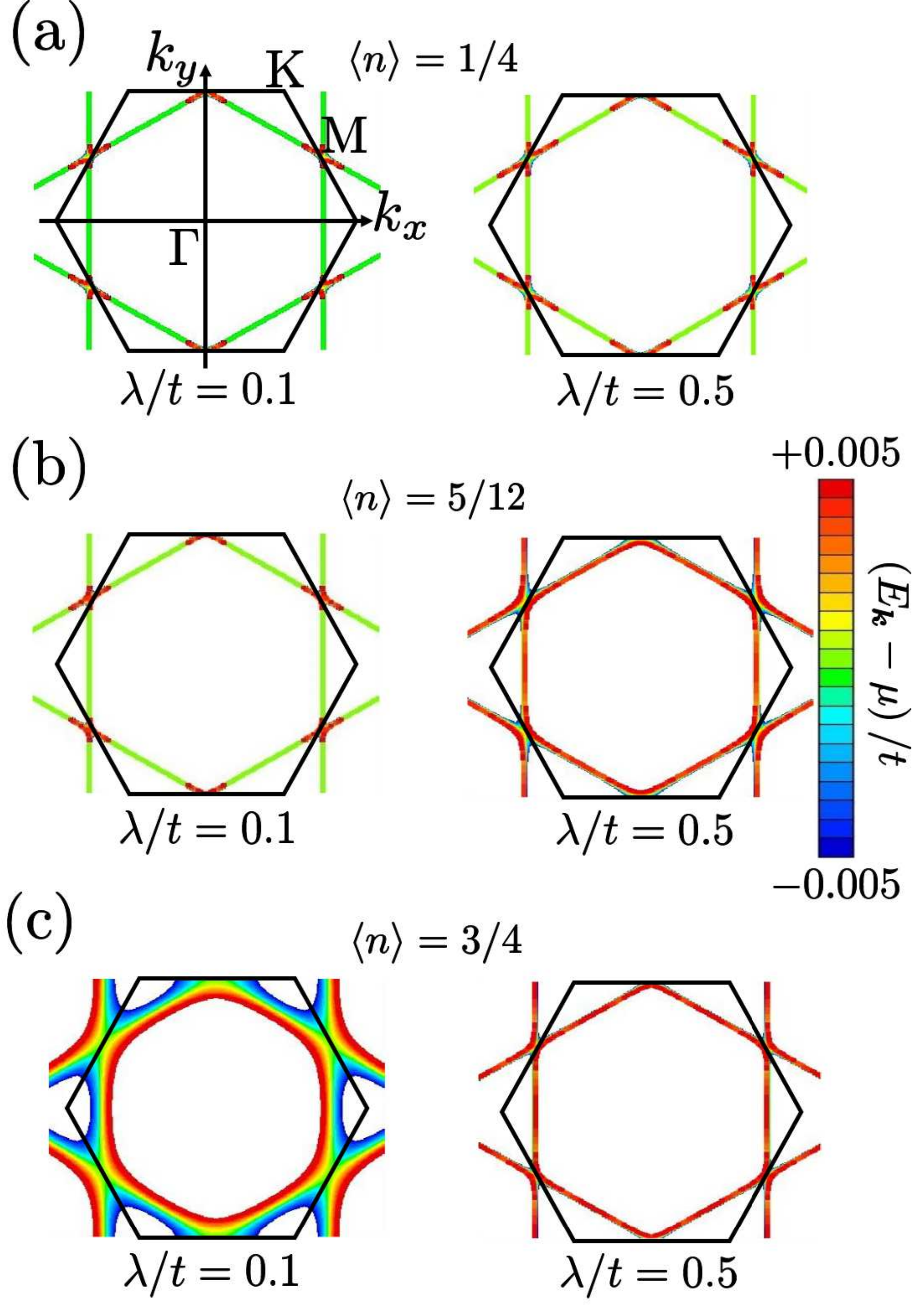}
		\caption{(Color online) Fermi surfaces and energy distributions around Fermi level at van Hove fillings:
			(a) $\langle n \rangle = 1/4$ ($\mu=\mu_1$), (b) $\langle n \rangle = 5/12$ ($\mu=\mu_2$), and (c) $\langle n \rangle = 3/4$ ($\mu=\mu_3$). 
			The black solid line represents the 1st Brillouin zone. 
			The color scale represents the energy distribution with $\left| E_{\bm{k}} - \mu \right| \le 0.005t$ around the Fermi level at van Hove fillings. }
		\label{FS}
	\end{center}
\end{figure}

Figure \ref{FS} shows Fermi surfaces and energy distributions within the small energy range ($\left| E_{\bm{k}} - \mu \right| \le 0.005t$) around the Fermi level at VH fillings $\left< n \right> = 1/4,5/12$, and $3/4$. 
The Fermi surfaces at all VH fillings have the hexagonal structure with the perfect nesting in the honeycomb bands even for $\lambda \ne 0$ and in the highest band except for $\lambda=0$.  
With increasing distance from VH fillings and/or the SOC, Fermi surfaces deviate from the hexagonal structure gradually. In particular, the deviation from the hexagonal structure in the upper honeycomb band is larger than that in the lower honeycomb band. It indicates that the number of states around the VH filling in the upper honeycomb band further increases with increasing SOC.

On the other hand, the introduction of the SOC leads to the hexagonal structure of the Fermi surface with the perfect nesting in the highest band. 
The distribution of the energy dispersion around the VH filling is enhanced with increasing $\lambda$ depicted in Fig. \ref{FS}(c), which indicates that the amplitude of the DOS decreases. 
However, compared with the cases at other VH fillings in the honeycomb bands,  there exists a larger DOS in the weak SOC region. Therefore, the appearance of the $d+id$ state in the highest band is expected for $\lambda \neq 0$ owing to the hexagonal Fermi surface and the large amplitude of DOS.

\subsection{Hall conductivity and Chern number}
\begin{figure}[b]
\begin{center}
\includegraphics[angle=90,width=6cm]{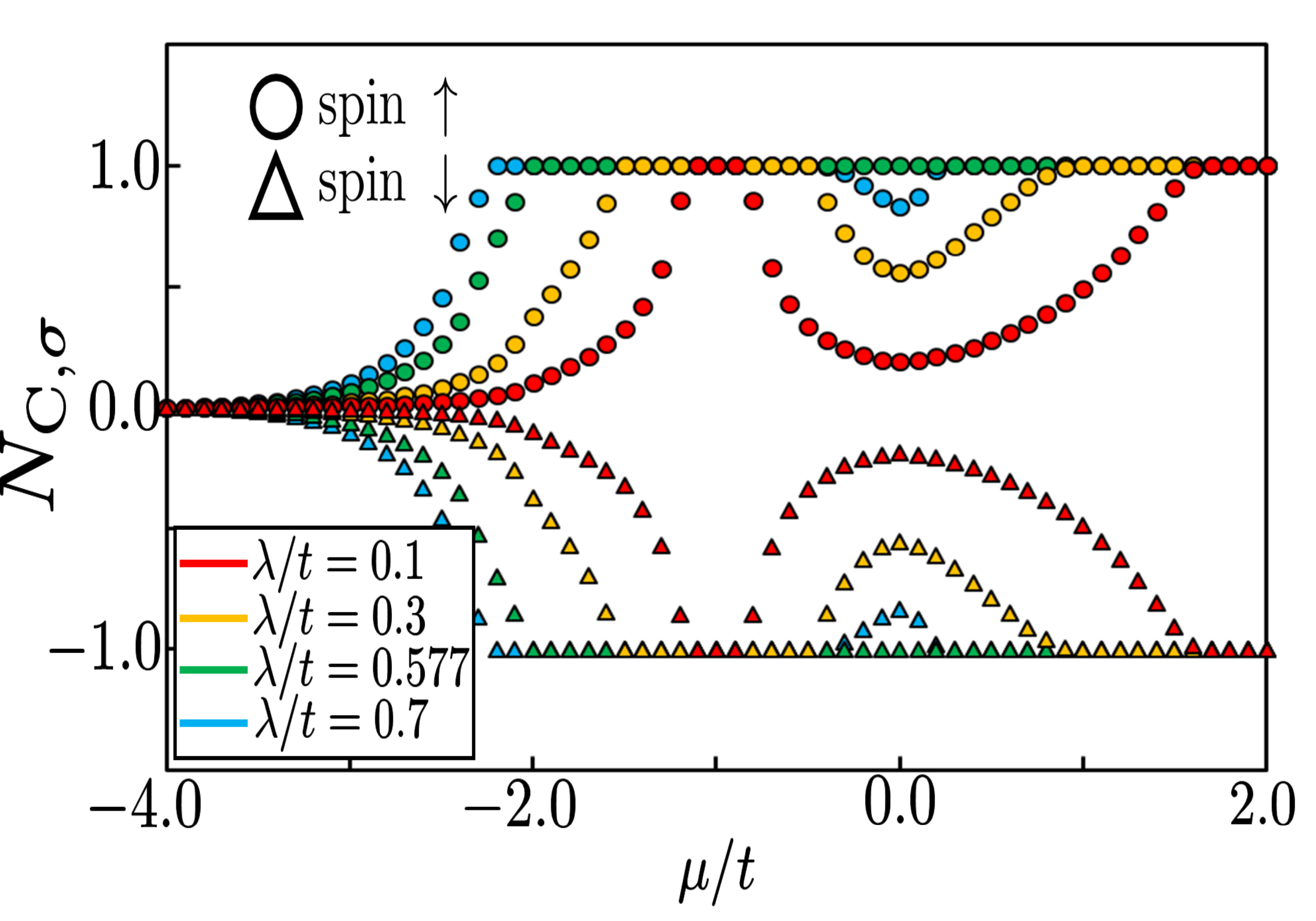}
\caption{(Color online) Spin-dependent Hall conductivity $\sigma_{xy}^{\uparrow (\downarrow)} (\mu)$ for several choices of $\lambda$. The corresponding spin $\sigma=\uparrow(\downarrow)$ is represented by circles (triangles).} 
\label{HC}
\end{center}
\end{figure}
In the honeycomb lattice, the SOC connecting NNN bonds induces the fully opened gap. The topologically nontrivial insulating phase appears when the chemical potential lies in the gap\cite{Kane_Mele,Kane_Mele2}. A similar topologically nontrivial insulating phase on the kagome lattice is also proposed by the introduction of the SOC connecting NNN bonds\cite{TI_kagome}. 
Thus, we examine the topological properties of the normal phase by calculating the Hall conductivity obtained from the Kubo formula\cite{TKNN,Kohmoto}, which is given by
\begin{align}
\label{Kubo_formula}
\sigma_{xy} = \frac{i}{\gamma N} \sum_{\bm{k},\alpha,\beta} \frac{\left< \alpha \right| \hat{j}_{\bm{k}}^x \left| \beta \right>\left< \beta \right| \hat{j}_{\bm{k}}^y \left| \alpha \right> }{E_{\bm{k}\alpha} - E_{\bm{k}\beta}} \frac{f(E_{\bm{k}\alpha})-f(E_{\bm{k}\beta})}{E_{\bm{k}\alpha} - E_{\bm{k}\beta}}, 
\end{align}
where $E_{\bm{k}\alpha}$ is the energy eigenvalue of $\hat{\xi}_\sigma (\bm{k})$ with the band index $\alpha$, and the factor $\gamma=\sqrt{3}/2$ denotes the area of the unit cell.
$f(\epsilon) \equiv 1/(\mathrm{e}^{\beta \epsilon} + 1)$ is the Fermi distribution function, and the current operator is defined as $\hat{j}^\mu_{\bm{k}} = \partial H(\bm{k})/\partial k_\mu$ $(\mu=x,y)$. 
Since the Hamiltonian in the absence of the attractive interaction with $H(\bm{k})=\hat{\xi}_{\uparrow}(\bm{k})\oplus\hat{\xi}_{\downarrow}(\bm{k})$ conserves the $z$-component of the spin, we calculate the spin-dependent Hall conductivity $\sigma_{xy}^{\uparrow(\downarrow)}$ by using the spin-block Hamiltonian $H_{\uparrow (\downarrow)} (\bm{k}) = \hat{\xi}_{\uparrow (\downarrow)} (\bm{k})$. 
Then, the spin-dependent Chern number is defined as
\begin{align}
N_{\mathrm{C},\uparrow(\downarrow)}=2\pi\sigma^{\uparrow(\downarrow)}_{xy}. 
\end{align}

Figure \ref{HC} shows the spin-dependent Hall conductivity on the kagome lattice as a function of the chemical potential $\mu$ for several choices of $\lambda$. 
When $\mu$ lies in the energy gap $\Delta_{\mathrm{I}}$ or $\Delta_{\mathrm{II}}$, the spin-dependent Chern number has an integer value with $N_{\mathrm{C},\sigma} =\mathrm{sgn}(\sigma)$. Although the total Chern number $N_{\mathrm{C}} (= N_{\mathrm{C},\uparrow} + N_{\mathrm{C},\downarrow})$ vanishes, the spin Chern number $N_{\mathrm{C}}^s$ defined as the difference $ (= N_{\mathrm{C},\uparrow} - N_{\mathrm{C},\downarrow} = +2)$ becomes a nonzero integer. This result indicates that the topologically nontrivial phase appears.
In addition to the topological state at $\mu$ in the gap $\Delta_{\mathrm{I}}$, which is essentially identical to that from the Kane-Mele model, there exists another topological state at $\mu/t\sim 2$ in the gap $\Delta_{\mathrm{II}}$, which is proper on the kagome lattice. 

\section{Superconducting Phase}
\subsection{Superconducting phase diagram}
\label{section_SCPD}
\begin{figure}[b]
\begin{center}
\includegraphics[angle=0,width=8.0cm]{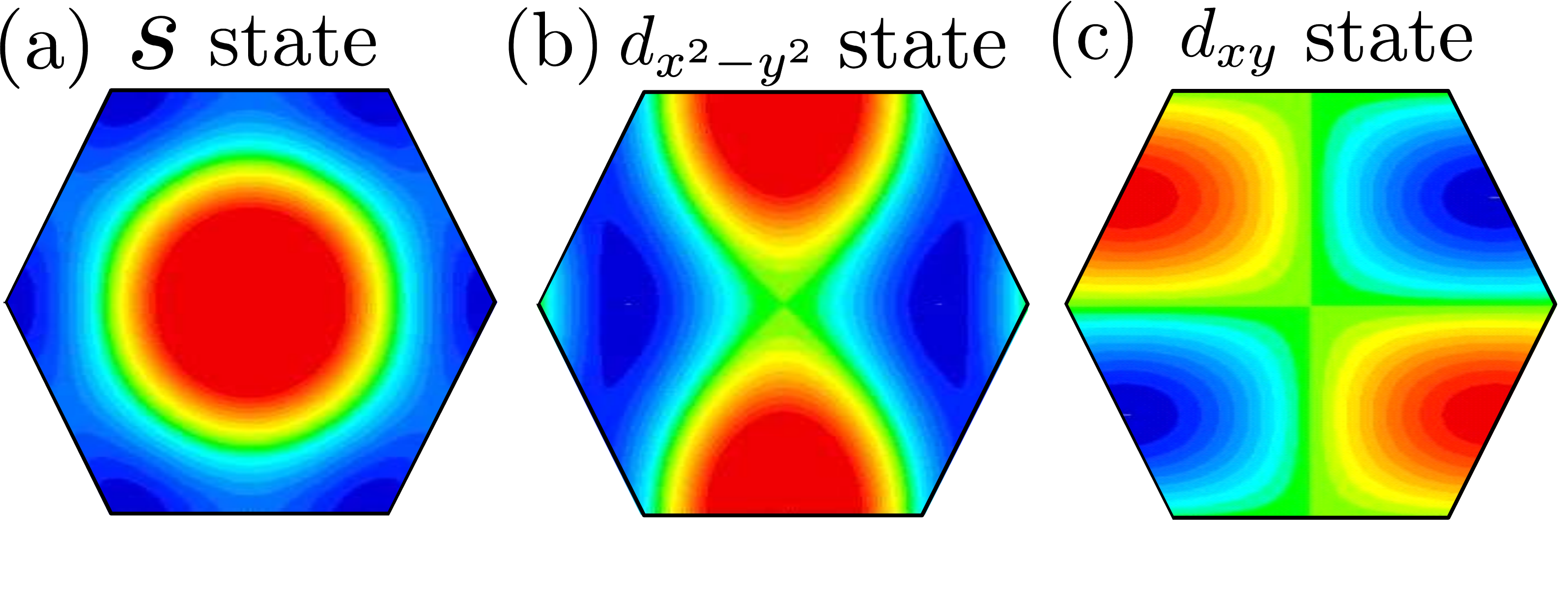}
\caption{(Color online) Gap functions for 3rd NN parings with positive (warm color) and negative (cold color) signs in the 1st Brillouin zone. The panels (a), (b), and (c) represent the $s$ state, $d_{x^2-y^2}$ state, and $d_{xy}$ state, respectively.  }
\label{order_parameters}
\end{center}
\end{figure}

Next, let us discuss the property in the superconducting phase at zero temperature in this subsection. 
Order parameters are obtained by solving the BdG Hamiltonian Eq. (\ref{BdG}) self-consistently. The most stable superconducting state is determined from the lowest-energy state in the obtained results by using several initial values of order parameters as follows:
\begin{align}
\label{initial_value}
\hat{\Delta}\equiv 
\begin{pmatrix}
\Delta_{M,1}\\
\Delta_{M,2}\\
\Delta_{M,3} 
\end{pmatrix}
= \left\{ \begin{array}{cc}
\frac{1}{\sqrt{3}}\left( 1 , 1 , 1 \right)^T & (s~\mathrm{state}) \\
\frac{1}{\sqrt{6}} \left(-1 , 2 , -1 \right)^T & (d_{x^2 - y^2}~\mathrm{state}) \\
\frac{1}{\sqrt{2}}\left( 1 , 0 , -1 \right)^T& (d_{xy}~\mathrm{state}) \\
\frac{1}{\sqrt{3}} \left( \mathrm{e}^{i\frac{2\pi}{3}} , 1 , \mathrm{e}^{i\frac{4\pi}{3}} \right)^T& (d+id~\mathrm{state})
\end{array} \right.,
\end{align}
where ``$s$ state'', ``$d_{x^2-y^2}$ state'', and ``$d_{xy}$ state'' represent the symmetries of the order parameter, which are defined as $\hat{\Delta}_s$, $\hat{\Delta}_{x^2- y^2}$, and $\hat{\Delta}_{xy}$, respectively.
The gap function is given by $\Delta_{M}(\bm{k}) = 2\sum_{i=1}^3 \Delta_{M,i} \mathrm{cos}\left( 2 \bm{k}\cdot \bm{a}_i \right)$. Figures \ref{order_parameters}(a)-(c) show gap functions in the Brillouin zone for each set. 

Since the upper three sets in Eq. (\ref{initial_value}) are orthogonal to each other and normalized, the order parameter obtained from solving the BdG equation self-consistently is represented as a linear combination of the sets as,
\begin{align}
\hat{\Delta}_M = \alpha_s \hat{\Delta}_s + \alpha_{x^2 - y^2} \hat{\Delta}_{x^2 - y^2} + \alpha_{xy} \hat{\Delta}_{xy}, 
\end{align}
with complex coefficients $( \alpha_s,\alpha_{x^2-y^2},\alpha_{xy})$ describing the superconducting state. For example, the pure $d_{x^2-y^2} + i d_{xy}$ state is represented by $( \alpha_s, \alpha_{x^2-y^2} , \alpha_{xy} ) \propto \left( 0 , 1, i\right)$. 
The Chern number $N_{\mathrm{C}}$ in the superconducting state is defined in a similar manner to the normal phase in Eq. (\ref{Kubo_formula})~\cite{imai13}. 

\begin{figure}[t]
	\begin{center}
		\begin{tabular}{c}
			\begin{minipage}{8.0cm}
			\begin{center}
			\includegraphics[width=8.0cm]{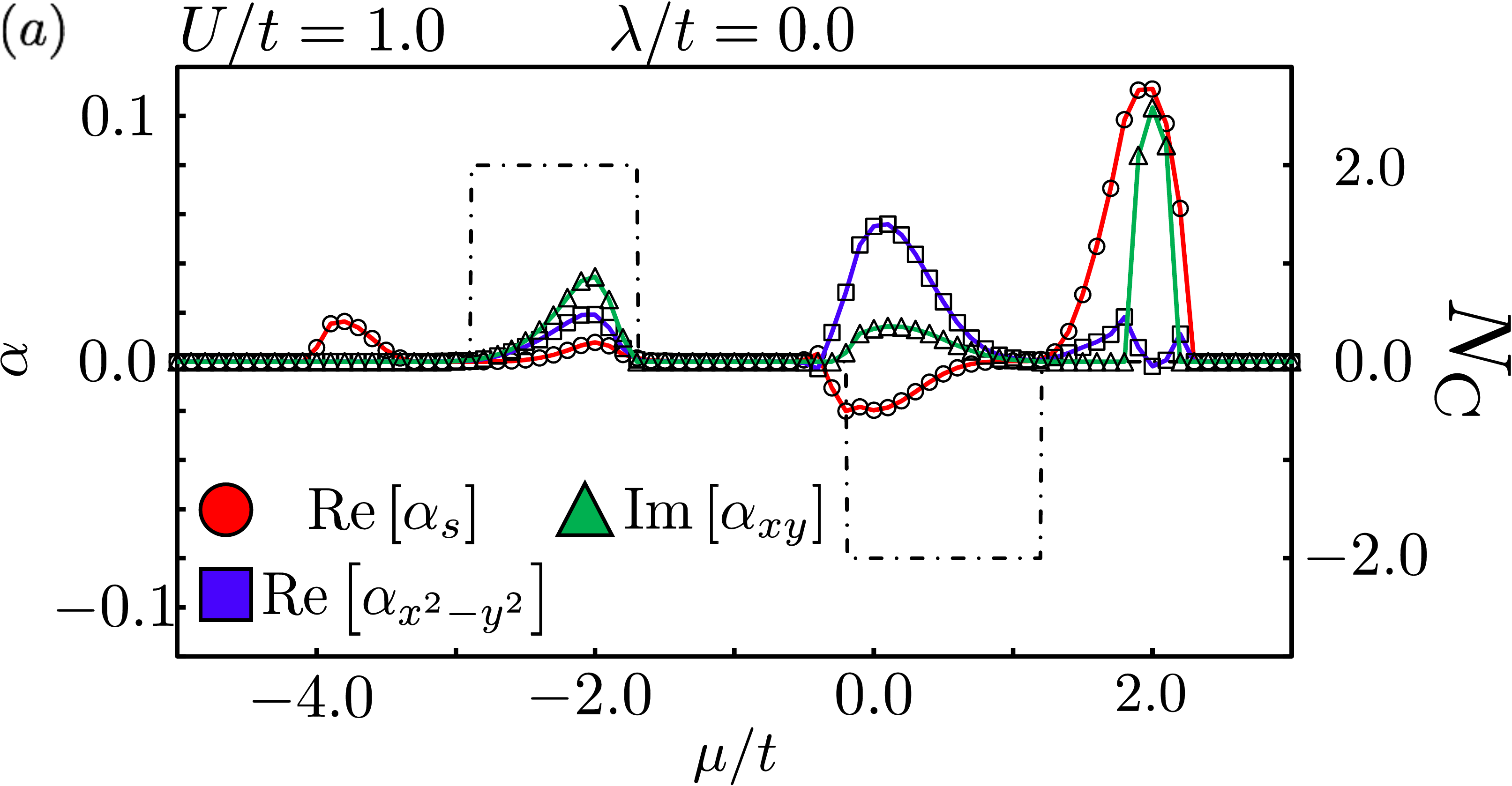}
			\end{center}
			\end{minipage}\\
			\vspace{0.5cm}\\
			\begin{minipage}{8.0cm}
			\begin{center}
			\includegraphics[width=8.0cm]{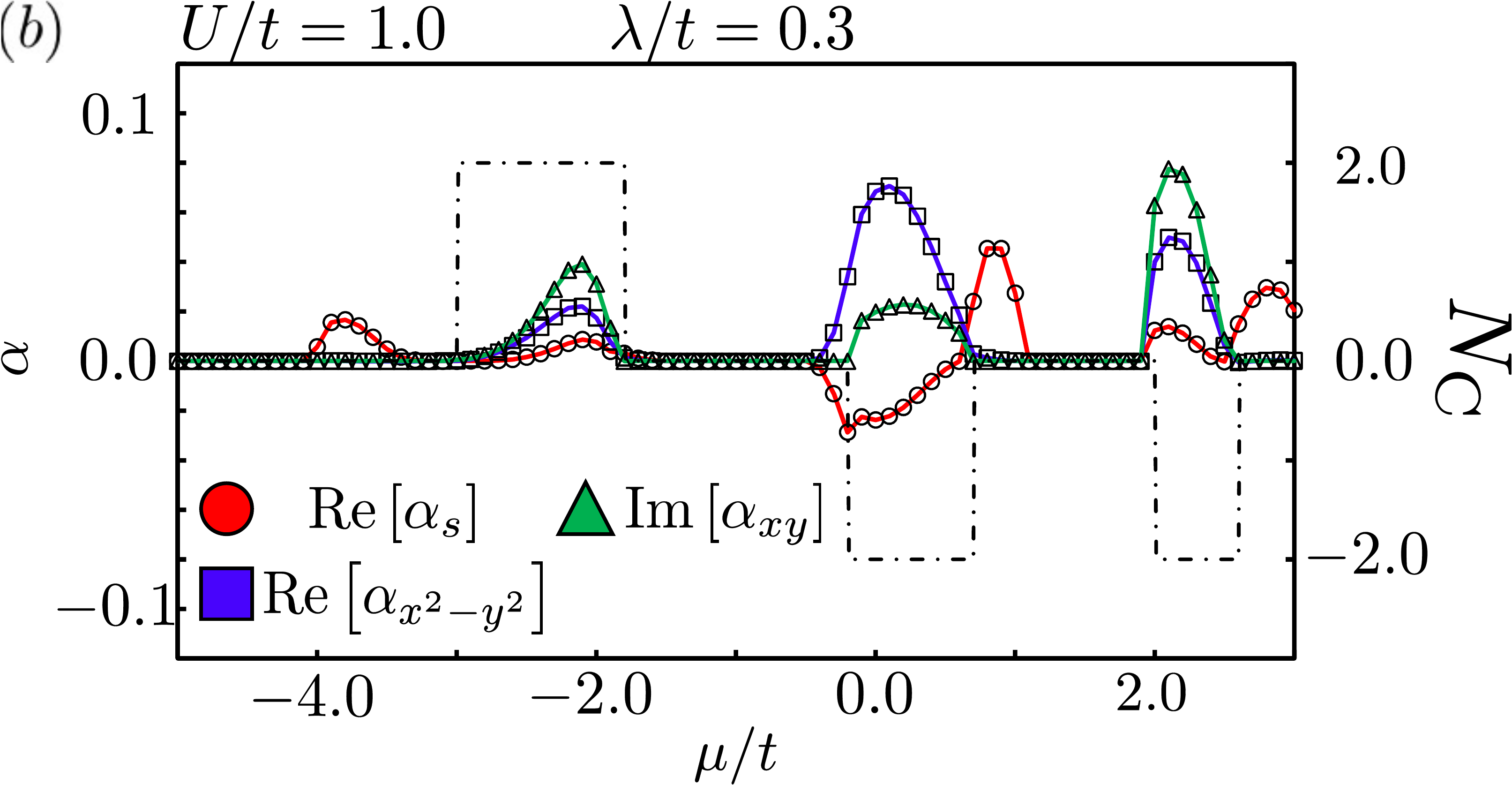}
			\end{center}
			\end{minipage}
		\end{tabular}
		\caption{(Color online) $\alpha$ (solid lines) and Chern number (dashed-dotted line) as functions of $\mu$ for $U/t=1$. The upper and lower panels represent $\lambda/t=0.0$ and $\lambda/t=0.3$, respectively. }
		\label{alpha}
	\end{center}
\end{figure}

Figure \ref{alpha} shows each $\alpha$ and the Chern number as functions of $\mu/t$ for $U/t=1$ in the most stable state. 
Note that other components ($\mathrm{Im}[\alpha_{s}]$, $\mathrm{Im}[\alpha_{x^2-y^2}]$, and $\mathrm{Re}[\alpha_{xy}]$) vanish in the whole parameter region and the obtained superconducting states consist of multiple components in the wide range of parameters.  

First, let us discuss the superconducting state without the SOC as shown in Fig. \ref{alpha}(a). 
In the honeycomb lattice model, the ``pure" $d+id$ states appear around VHSs where the amplitude of $\alpha_{x^2 -y^2}$ is identical to that of $\alpha_{xy}$ in the weak-coupling region. 
On the other hand, all components of $\alpha$ have finite values around $\mu=\mu_1(=-2t)$ and $\mu=\mu_2(=0)$ on the kagome lattice owing to the lower lattice symmetry in comparison with the honeycomb lattice, where $\left|\alpha_{x^2 -y^2}\right|$ is not equal to $\left| \alpha_{xy}\right|$ and the $s$ state component becomes finite ($\alpha_s\ne 0$).  

This inequivalence between $|\alpha_{x^2-y^2}|$ and $|\alpha_{xy}|$ results from the inequivalence among three different 3rd NN pairings on the kagome lattice depicted in Fig. \ref{fig_model}(b). 
Although the $s$ component appears in addition to the $d_{x^2 -y^2}$ and $d_{xy}$ components around VHSs in the honeycomb bands, the Chern number still remains finite ($N_C=\pm 2$), which is defined as the ``$d_1+id_2$" state in this study. 
The $d_1$ and $d_2$ states stand for the $d_{x^2-y^2}$ and $d_{xy}$ ones, respectively. 
$\mathrm{Im}[\alpha_{xy}]$ vanishes at $1.2<\mu/t<1.8$ and $\mathrm{Re}[\alpha_{s}]$ is dominant, which is defined as the topologically trivial ``$s+d_1$'' state with TRS. 
Around $\mu=\mu_3(=2t)$, there exist $\mathrm{Re}[\alpha_{s}]$ and $\mathrm{Im}[\alpha_{xy}]$ components with a very small amplitude of $\mathrm{Re}[\alpha_{x^2-y^2}]$, which is defined as the topologically trivial ``$s+id_2$'' state with TRS breaking. The amplitude of the Chern number distinguishes the $d_1+id_2$ and $s+id_2$ states.  

We now turn on the SOC, shown in Fig. \ref{alpha}(b). In the case of $\mu/t\lesssim 0.8$, the amplitudes of $\alpha$ are similar to those without the SOC. 
Since the SOC generates the energy gap $\Delta_{\mathrm{II}}(\approx 1.04t \text{ for }\lambda=0.3t)$ between the upper honeycomb band and the highest band, the top of the upper honeycomb band shifts to the low-energy region. Thus, the $s+d_1$ state also shifts to the top of the upper honeycomb band ($\mu/t\sim 1$). 

On the other hand, around $\mu=\mu_3$, the amplitude of $\mathrm{Re}[\alpha_{s}]$ is reduced and that of $\mathrm{Re}[\alpha_{x^2-y^2}]$ is enhanced, so that the $d_1+id_2$ state with the finite Chern number appears. 
Since the Fermi surface in the highest band has hexagonal symmetry due to the SOC, the strong nesting effect leads to the $d_1+id_2$ state. 
This result indicates that the SOC gives rise to the topological superconducting state in the highest band on the kagome lattice. 

\begin{figure}[t]
\begin{center}
\includegraphics[width=7cm]{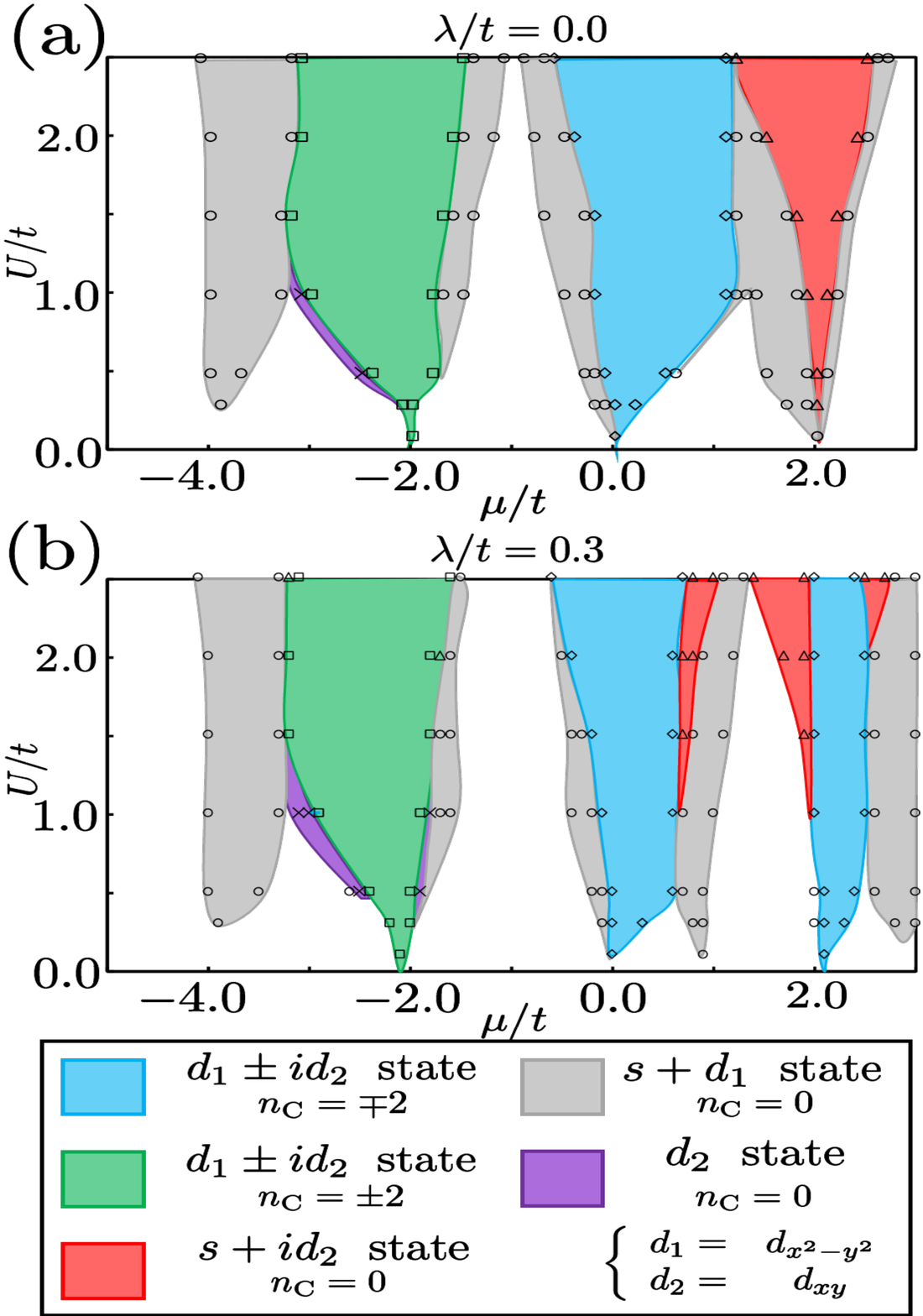}
\caption{(Color online) Phase diagram of the superconducting state on kagome lattice. The upper and lower panels represent $\lambda/t=0.0$ and $\lambda/t=0.3$, respectively. 
}
\label{SCPD}
\end{center}
\vspace{-0.8cm}
\end{figure}

Figure \ref{SCPD} shows the phase diagram in the most stable superconducting state without/with the SOC where the superconducting states are determined by the amplitude of $\alpha$ and the Chern number in a similar manner shown in Fig. \ref{alpha}. 
The $d_1\pm id_2$ states have the finite Chern number ($N_{\mathrm{C}} = \pm 2$) around VHSs, and other superconducting states are topologically trivial. 

In the absence of the SOC, the component of the $d_1\pm id_2$ state is dominant around $\mu=\mu_1$ and $\mu=\mu_2$ in the honeycomb bands. 
On the other hand, the energy dispersions near the bottom of the lower honeycomb band and the top of the upper honeycomb band are symmetric with respect to $\varepsilon=0$ for $\lambda=0, \mu/t=-1$, and there are circular Fermi surfaces around the $\Gamma$ point ($\mu/t\simeq -4$ and $2$). 
We stress that there exists the topologically trivial $s+i d_2$ state with the broken time-reversal symmetry around $\mu = \mu_3$ mainly in the flat band and may be difficult to appear on the honeycomb lattice. 

In the presence of the SOC ($\lambda/t=0.3$), the phase diagram does not change markedly in the honeycomb bands qualitatively. 
On the other hand, the $s+i d_2$ with $N_{\mathrm{C}} =0$ state found for $\lambda/t=0$ around $\mu=\mu_3$ turns into the $d_1 \pm id_2$ state with  $N_{\mathrm{C}} =\mp2$ in the wide range of parameters, which is attributed to the Fermi surface nesting with the hexagonal symmetry induced by the SOC in the highest band. 

The SOC leads to the hexagonal Fermi surface with perfect nesting at $\mu=\mu_3$, where the sufficiently small $U$ generates the $d_1+id_2$ state and further increase in $U$ generates the $s+id_2$ state. Since the phase boundary between the $d_1+id_2$ and $s+id_2$ states is determined by the competition between the width of the highest band and the amplitude of $U$, the parameter region of the appearance of the $d_1+id_2$ state is enlarged around $\mu=\mu_3$ with increasing amplitude of the SOC. 

Note that in the large $U$ region ($U/t>1$), although the order parameter changes continuously around $\mu/t=2$, the Chern number switches at $\mu/t=2$. It is attributed to the change in the Fermi surface topology where the Fermi surface vanishes for $\mu/t<2$. 

\begin{figure}[t]
\begin{center}
\begin{tabular}{cc}
\begin{minipage}{4.3cm}
\begin{center}
\includegraphics[angle=90,width=4.3cm]{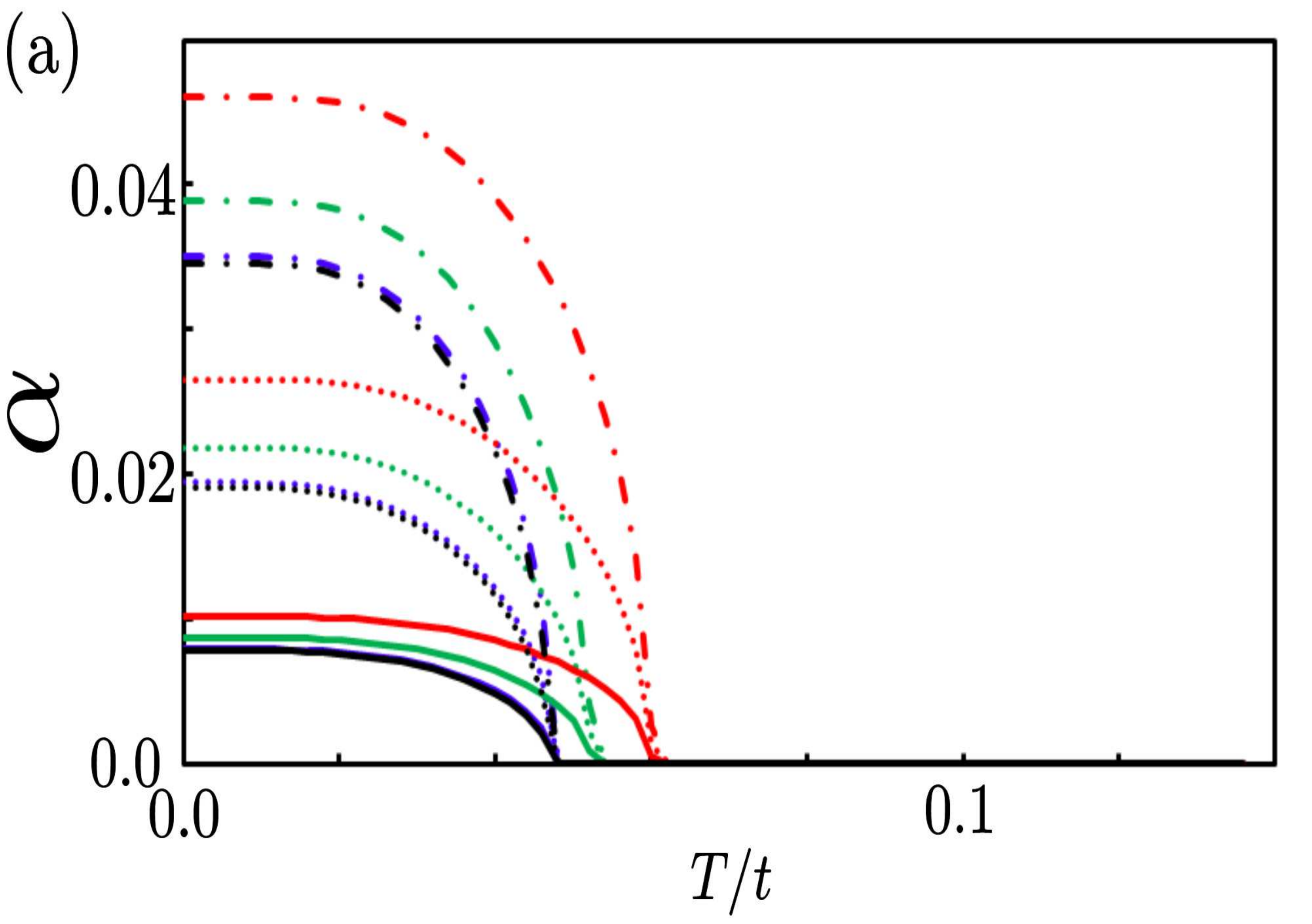}
\end{center}
\end{minipage}
\begin{minipage}{4.3cm}
\begin{center}
\includegraphics[angle=90,width=4.3cm]{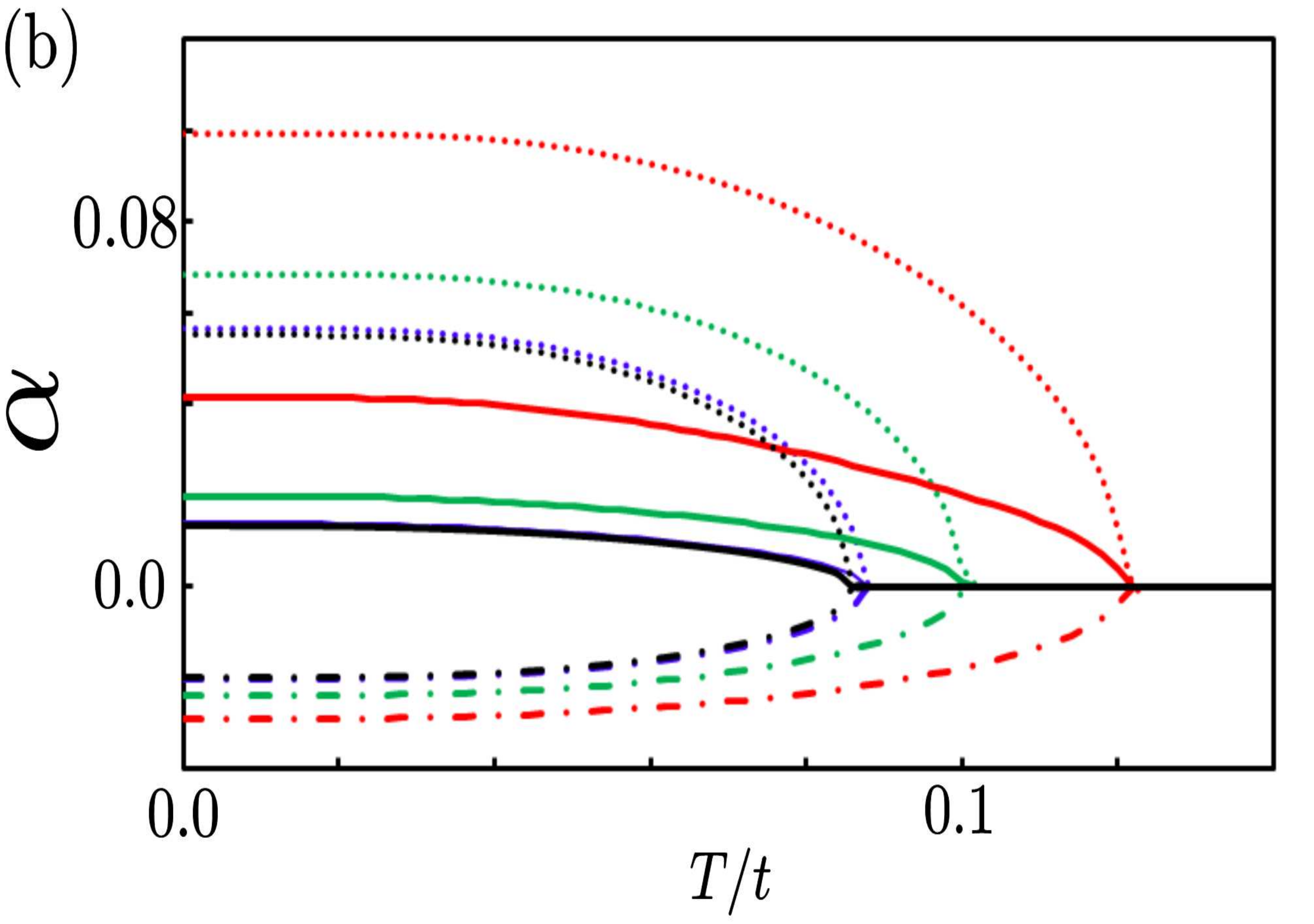}
\end{center}
\end{minipage}\\
\begin{minipage}{4.3cm}
\begin{center}
\includegraphics[angle=90,width=4.3cm]{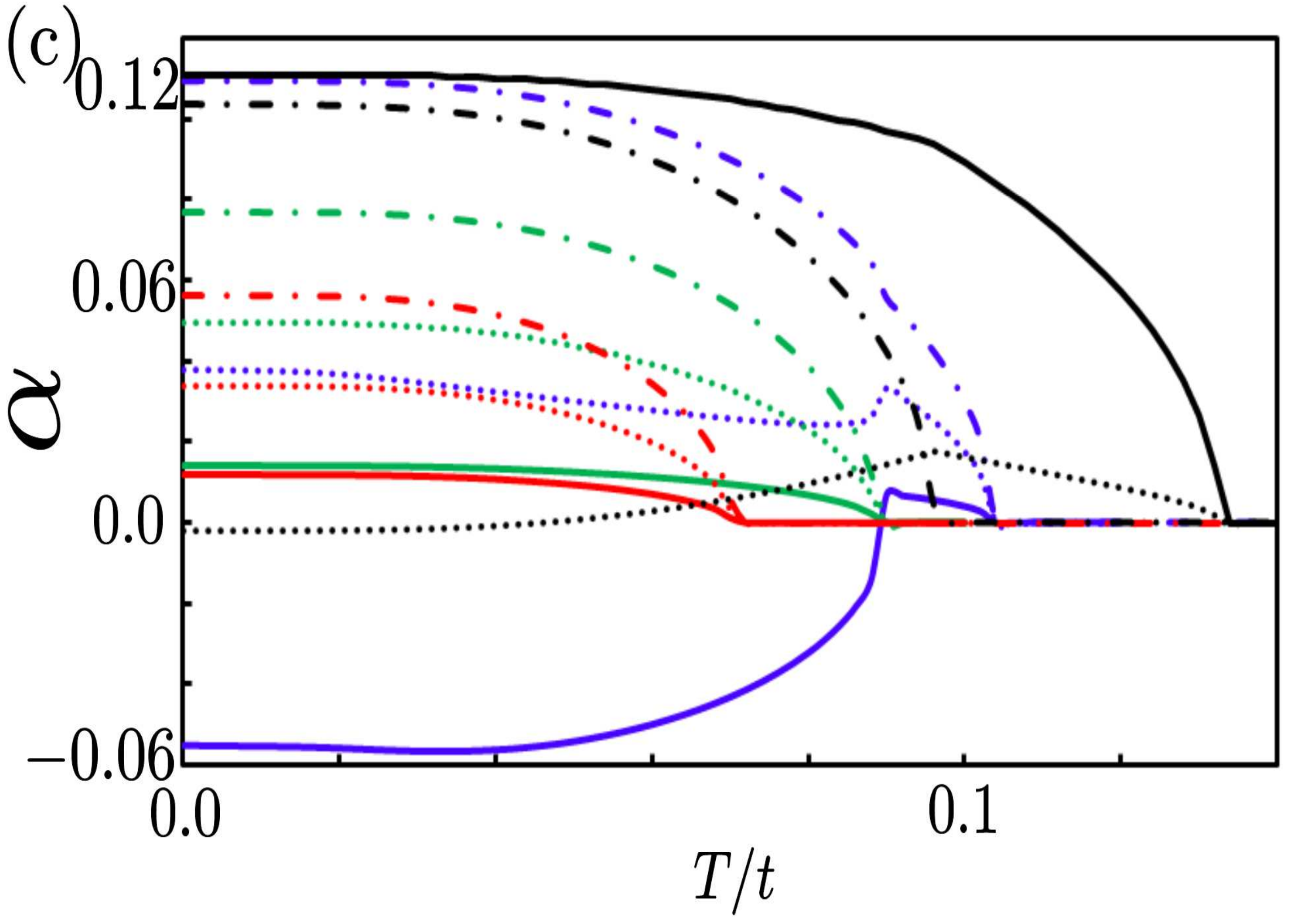}
\end{center}
\end{minipage}
\begin{minipage}{4.3cm}
\begin{center}
\includegraphics[angle=90,width=4.3cm]{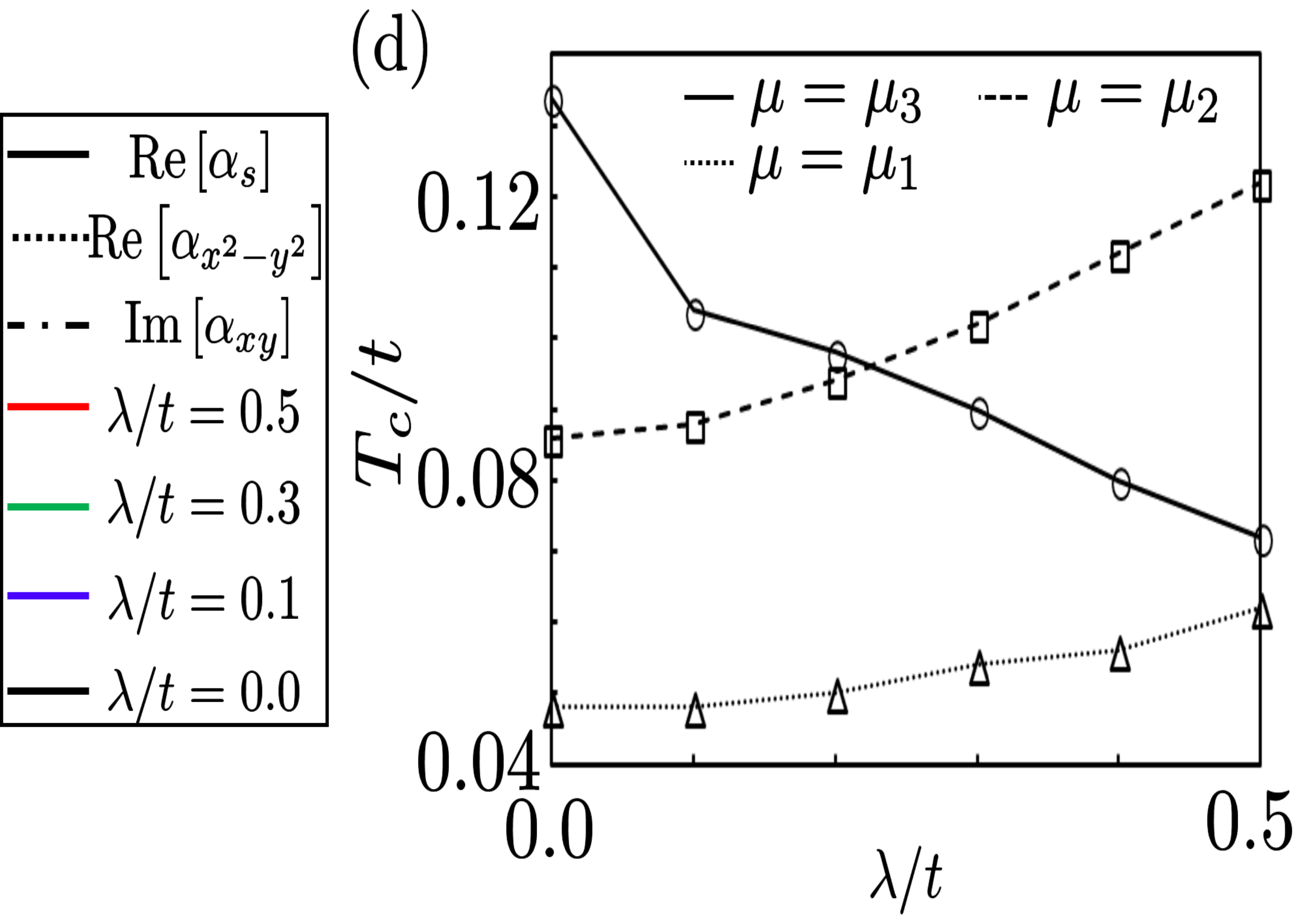}
\end{center}
\end{minipage}
\end{tabular}
\caption{(Color online) (a)-(c): Temperature dependences of order parameters at van Hove fillings (a) $\mu=\mu_1$, (b) $\mu=\mu_2$, and (c) $\mu=\mu_3$ for $U/t=1$. (d): Transition temperature ($T_c$) plotted as a function of $\lambda$. }
\label{Temperature_dependence}
\end{center}
\vspace{-0.8cm}
\end{figure}

Next, we discuss the temperature dependences of order parameters for $U/t=1$ at VHSs, as plotted in Figs. \ref{Temperature_dependence}(a)--(c). 
The superconducting transition temperature $T_c$ is also affected by the SOC. 
With increasing $\lambda$, the order parameters increase monotonically at $\mu=\mu_1$ and $\mu=\mu_2$ in the whole temperature region and the transition temperatures also increase. Note that the sign of $\text{Re}[\alpha_{x^2-y^2}]/\text{Im}[\alpha_{xy}]$ is directly associated with the sign of the Chern number in the $d_1\pm id_2$ state. 
On the other hand, the order parameter decreases generally with increasing $\lambda$ at $\mu=\mu_3$. 
The amplitude of the order parameter is larger than those at $\mu_1$ and $\mu_2$ since there exists a large amplitude of the DOS around $\mu=\mu_3$. 
Thus, $T_c$ at $\mu=\mu_3$ is higher than those at $\mu=\mu_1$ and $\mu=\mu_2$ for a small $\lambda$, as depicted in Fig. \ref{Temperature_dependence}(d). 
These results indicate that the $d_1+id_2$ state in the highest band on the kagome lattice may have a large transition temperature in comparison with that on the honeycomb lattice. 

Note that the nonmonotonic behavior appears with increasing temperature at $\mu=\mu_3$ for the small-$\lambda$ region. Around $T=T_c$, while the $d_1+id_2$ component is dominant, the amplitude of order parameters changes discontinuously and the $s$ component is enhanced in the low-temperature region for $\lambda/t=0.1$. However, the Chern number still becomes a finite integer in the low-temperature region, which indicates that the superconducting state still maintains the $d_1+id_2$ state.  

However, for the rather small $\lambda~(\lesssim 0.09t)$ while the $s$ component is further enhanced and the $s+id_2$ state is realized in the low-temperature region, the $d_2$ component vanishes and  the $s+d_1$ state remains around $T=T_c$. It results from the above-mentioned inequivalence between the $d_1$ and $d_2$ states.
When the amplitude of the effective attractive interaction ($\sim U \left|\alpha\right|$) is smaller than the width of the highest band ($W=2\sqrt{3} \lambda$) around $T=T_c$, the effect of Fermi surface nesting plays an important role in giving rise to the $d_1+i d_2$ state.

\subsection{Thermal Hall conductivity}
\label{Thermal_Hall_conductivity}
\begin{figure}[t]
\begin{center}
 \begin{tabular}{c}
  \begin{minipage}{7.0cm}
   \begin{center}
    \includegraphics[width=3.9cm,height=7.0cm,angle=90]{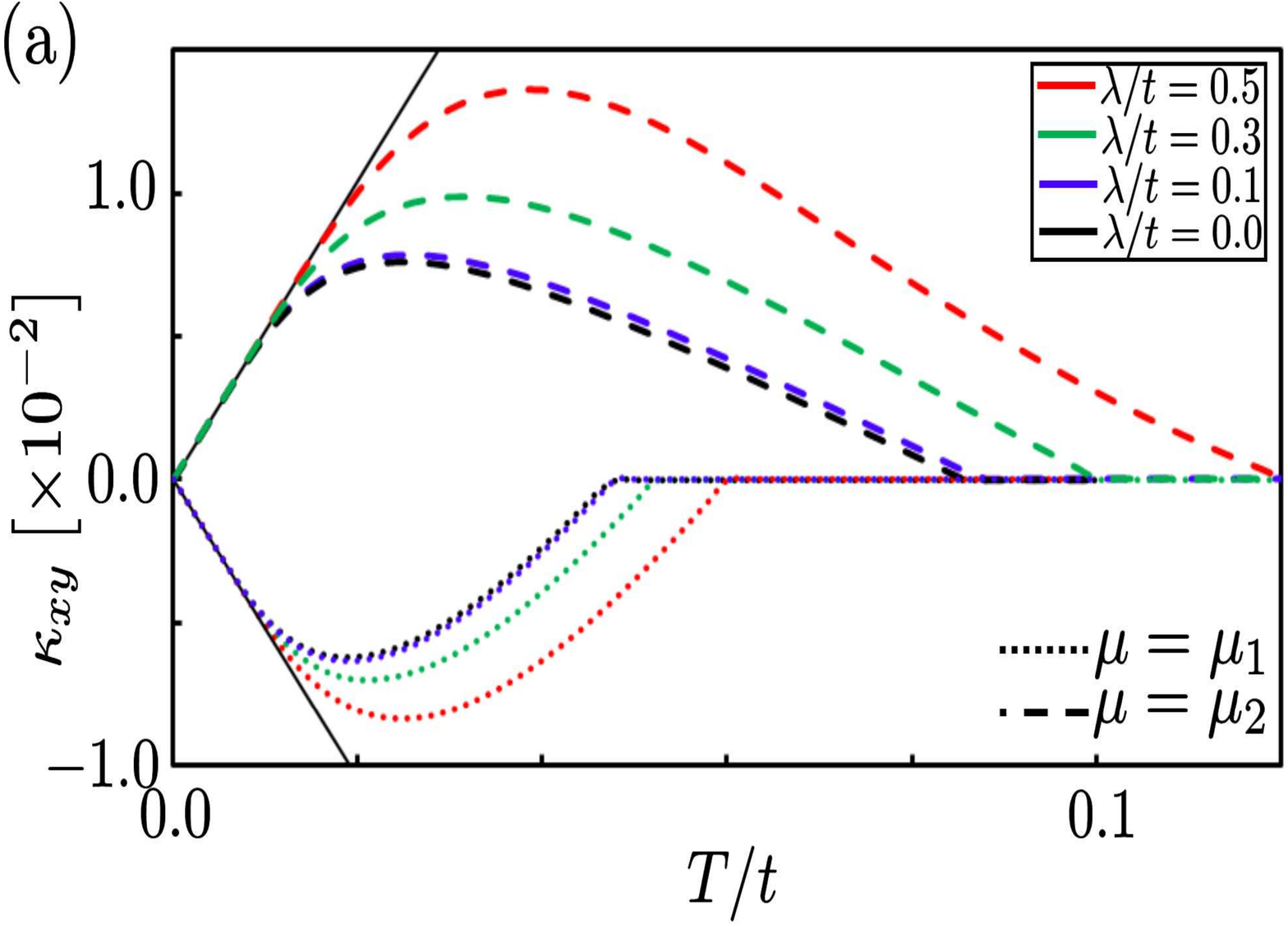}
   \end{center}
  \end{minipage}\\
  \begin{minipage}{7.0cm}
   \begin{center}
    \includegraphics[width=3.9cm,height=7.0cm,angle=90]{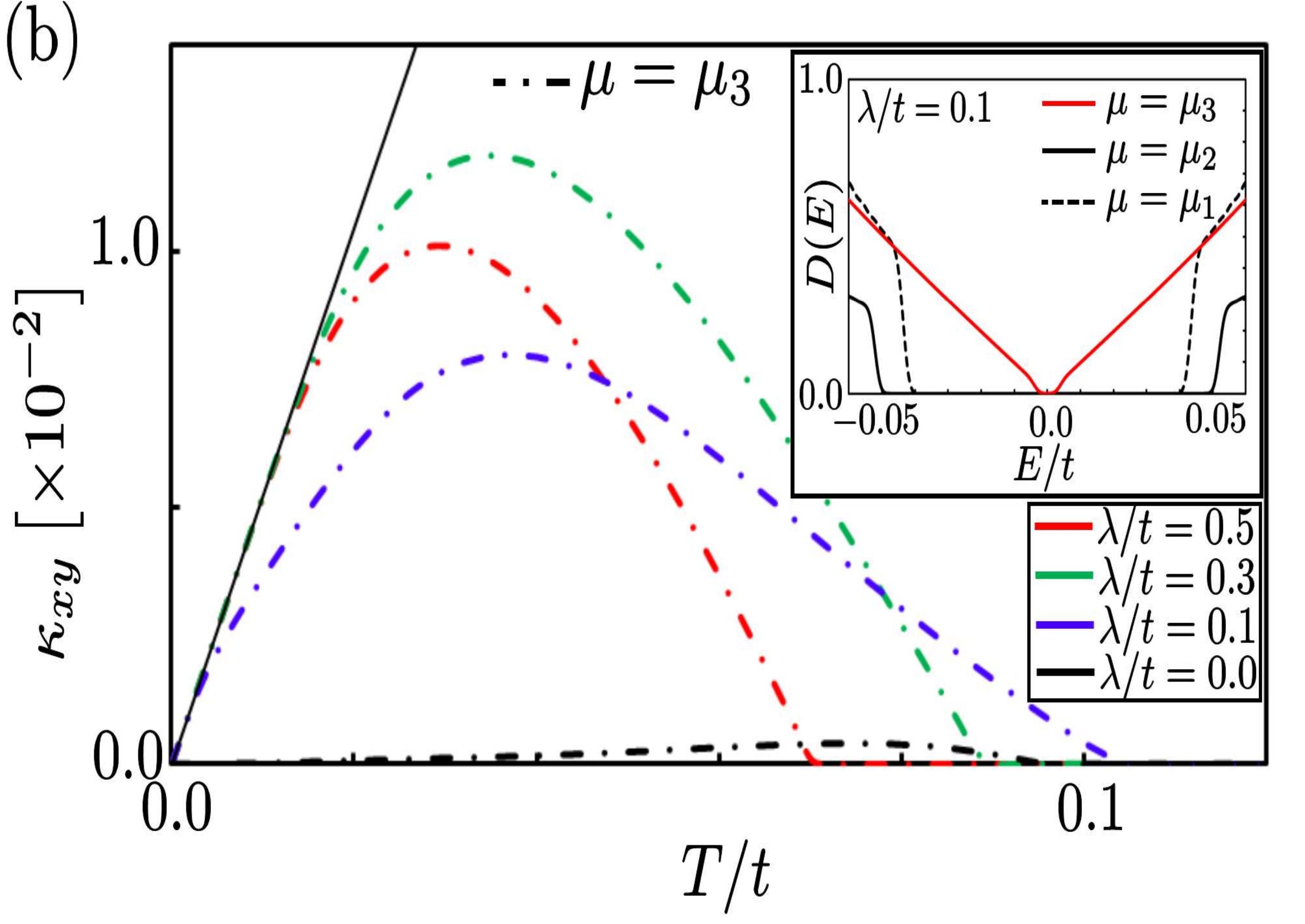}
   \end{center}
  \end{minipage}
 \end{tabular}
\caption{(Color online) Thermal Hall conductivity is plotted as a function of the temperature for several choices of $\lambda$. The black solid line represents the temperature-linear behavior given by Eq. (\ref{THC_tp}). (a) $\mu=\mu_1$ (dotted line) and $\mu=\mu_2$ (dashed line). (b) $\mu=\mu_3$ (dot-dashed line). The inset shows the DOS in the low-energy region at VHSs with $\lambda/t=0.1$ at $T=0$.}
\label{THC_sc}
\end{center}
\vspace{-0.8cm}
\end{figure}

The measurement of the thermal Hall conductivity gives information on the topological properties in TRS broken superconductors\cite{Thermal_Hall_conductivity}, which may enable us to distinguish superconducting states on the kagome lattice. 
It can be expressed as
\begin{align}
\label{THC}
\kappa_{xy} = -\frac{1}{TN} \sum_{\bm{k},\alpha} \mathrm{Im}\left[ \left< \frac{\partial u_{\bm{k}\alpha}}{\partial k_x} \bigg| \frac{\partial u_{\bm{k}\alpha}}{\partial k_y} \right> \right] \int dE E^2 \theta (E-E_{\bm{k}\alpha}) f'(E), 
\end{align}
where $f'(E)$ is the derivative of the Fermi distribution function. $T$ and $N$ denote the temperature and number of sites, respectively\cite{Thermal_Hall_conductivity}. $u_{\bm{k}\alpha}$ is the periodic part of the Bloch wave function of the BdG Hamiltonian for the wave vector $\bm{k}$ and the band index $\alpha$. $\mathrm{Im}\left[ \left< \partial_{k_x} u_{\bm{k}\alpha} | \partial_{k_y} u_{\bm{k}\alpha} \right> \right]$ is proportional to the $z$ component of the Berry curvature $\Omega(\bm{k})$. 

Moreover, in the low-temperature limit, the thermal Hall conductivity (Eq. (\ref{THC})) is proportional to temperature  and the Chern number\cite{Thermal_Hall_conductivity}, which is given by
\begin{align}
\label{THC_tp}
\kappa_{xy} \simeq -\frac{\pi}{12}N_{C} T. 
\end{align}
Therefore, in the low-temperature region, $\kappa_{xy}$ is uniquely determined by the topological properties of the superconducting state.  

The breaking of the TRS and inversion symmetry in the superconducting state gives rise to the finite thermal Hall conductivity.  
Therefore, $\kappa_{xy}$ will vanish in the $s+d_1$ state with the TRS. Although the $s+id_2$ state breaks the TRS and inversion symmetry, $\kappa_{xy}$ also vanishes because the $s+id_2$ state is the topologically trivial one. 
On the other hand, the $d_1+id_2$ state with the finite Chern number gives a finite $\kappa_{xy}$. 

Figures \ref{THC_sc}(a) and \ref{THC_sc}(b) show the temperature dependence of the thermal Hall conductivity for several combinations of $\left(\mu,\lambda\right)$ with $U/t=1.0$. 
For $\mu=\mu_1$, $\mu_2$, and $\mu_3$ with $\lambda/t \ge 0.3$, the $d_1+id_2$ states appear and the thermal Hall conductivity has a linear behavior with respect to temperature in the low-temperature region, whose slope corresponds to the Chern number. 

We can also confirm the temperature linear behavior even at $\mu=\mu_3$ at $\lambda/t=0.1$, but that occurs only in the temperature region lower than that at $\mu=\mu_1$ and $\mu=\mu_2$. The deviation from Eq. (\ref{THC_tp}) is well described by the exponential term $\mathrm{e}^{-\beta \Delta}$ with the quasiparticle gap $\Delta$\cite{Imai}. DOSs in the $d_1+id_2$ state at VHSs at absolute zero are shown in the inset in Fig. \ref{THC_sc}(b). The amplitude of the quasiparticle gap at $\mu=\mu_3$ is smaller than those at $\mu=\mu_1$ and $\mu=\mu_2$. It results from the mixing between the $s$ and $d_1$ components in the $d_1 +id_2$ state for $\mu=\mu_3$ and $\lambda=0.1t$. The $s$ state with the 3rd NN pairing has a circular line node unlike the local pairing. Therefore, when $\left| \alpha_s \right|$ becomes comparable to $\left| \alpha_{x^2-y^2} \right|$, the $s+d_1$ state corresponding to the real part of the $d_1 +id_2$ state can have point nodes at $k_x = 0$ and $k_y = 0$,  where the gap function describing the $d_2$ state corresponding to the imaginary part vanishes. Thus, the superconducting gap is strongly suppressed around $\mu=\mu_3$. 

In addition, the thermal Hall conductivity at $\mu=\mu_3$ for $\lambda=0$ shows a different temperature dependence from those for other choices of $\lambda$ in Fig. \ref{THC_sc}(b). In the low-temperature region, since the topologically trivial $s+i d_2$ component with broken TRS is dominant, $\kappa_{xy}$ vanishes. However, the $d_1$ component increases with increasing temperature, so that $\kappa_{xy}$ has a finite amplitude. On the other hand, the $d_2$ component vanishes at $T/t\sim 0.1$ and the $s+d_1$ component with the TRS becomes dominant at $0.1t<T<T_c\,(\sim 0.13t)$. Therefore, $\kappa_{xy}$ vanishes at the temperature lower than $T_c$.

\begin{figure}[t]
\begin{center}
\includegraphics[width=5.0cm,height=7.0cm,angle=90]{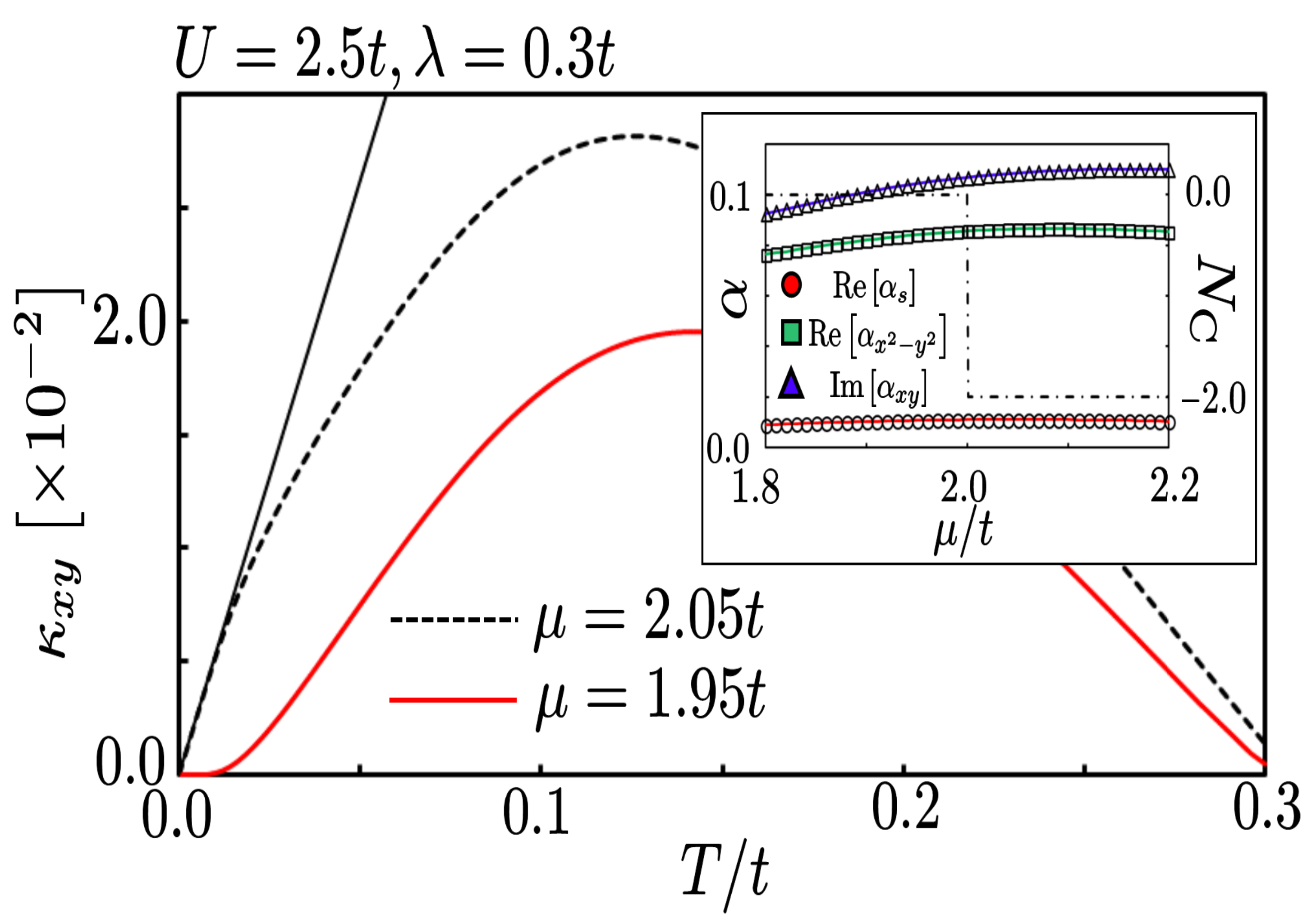}
\caption{(Color online) Thermal Hall conductivity as a function of temperature for $U/t=2.5$ and $\lambda/t=0.3$ around $\mu/t=2$. The inset shows order parameters around $\mu/t=2$ and the Chern number.}
\label{Lifshitz_transition}
\end{center}
\vspace{-0.8cm}
\end{figure}

Note that in the large-$U$ region, although the order parameters change continuously around $\mu/t=2$, the topological property switches, which is discussed in the previous subsection. 
The thermal Hall conductivity is strongly sensitive to the amplitude of the chemical potential, depicted in Fig. \ref{Lifshitz_transition}. $\kappa_{xy}$ has the temperature linear behavior and vanishes in the low temperature limit for $\mu/t>2$ and $\mu/t<2$, respectively, owing to the switch of the topological property. 

While the low-temperature behavior is strongly affected by the amplitude of $\mu$, the high-temperature one has a similar form with respect to the temperature around $\mu/t=2$.  Since the order parameters for $\mu/t=1.95$ and $2.05$ are almost the same, it is difficult to distinguish the superconducting states for $\mu/t \gtrsim 2$ and $\mu/t \lesssim 2$ in the high-temperature region. 

Here we demonstrate the large-$U$ and $\lambda$ case. However, since the switch of the temperature dependence of $\kappa_{xy}$ around $\mu/t=2$ is associated with the ratio $U/\lambda$, a similar behavior can occur even in the low-$U$ and $\lambda$ region. 

\section{Conclusions}
We have investigated the spin-singlet superconducting state and the temperature dependence of the thermal Hall conductivity  on the kagome lattice by the self-consistent BdG approach. In particular, we focus on the highest band and discuss the interplay between the amplitude of the attractive interaction and the SOC effect. 

In the normal phase, there exist perfectly hexagonal Fermi surfaces at all VH fillings in the honeycomb bands. The presence of the SOC leads to the dispersive highest band with the hexagonal symmetry, and also generates band gaps, where topological nontrivial states appear where that around $\mu/t=2$ is proper on the kagome lattice.   

In the superconducting phase, order parameters and the phase diagram are obtained. In the honeycomb bands, the $d_1+id_2$ states appear in the vicinity of VHSs. Since the amplitude of the DOS at VH fillings is enhanced by the SOC, $T_c$ increases with increasing $\lambda$. In the highest band around $\mu=\mu_3$, the $s+i d_2$ state appears for $\lambda = 0$. Since the Fermi surface in the highest band has the perfect nesting property induced by the SOC at $\mu=\mu_3$, the $d_1+id_2$ state appears for $\lambda>0$. When the width of the highest band depending on the amplitude of the SOC is comparable to the amplitude of the attractive interaction and/or temperature, the $s+id_2$ ($d_1+id_2$) character is dominant in the low-temperature region (around $T_c$). 

Moreover, the thermal Hall conductivity is proportional to the temperature with a prefactor that is uniquely related to the Chern number. While the $d_1+id_2$ state has the temperature linear behavior, the thermal Hall conductivity in the $s+i d_2$ state increases slightly with temperature owing to the increase in the $d_1$ component. In particular, the character of pairing symmetries changes in the highest band owing to the competition between the attractive interaction and the bandwidth, which may be detectable in the thermal Hall conductivity. Therefore, the measurement of the thermal Hall conductivity may enable us to distinguish the superconducting states on the kagome lattice. 
\section*{Acknowledgment}
We acknowledge valuable discussions with J. Goryo and S. Hoshino.

\end{document}